\documentclass[fleqn,usenatbib]{mnras}

\usepackage{newtxtext,newtxmath}
\usepackage{longtable}
\usepackage{booktabs}

\usepackage[T1]{fontenc}

\usepackage{setspace}

\usepackage{graphicx}	
\usepackage{amsmath}	

\usepackage{xcolor}
\definecolor{pink}{rgb}{0.98, 0.38, 0.5}

\newcommand{\Vr}{$V_r$}
\newcommand{\Vphi}{$V_{\phi}$}

\newcommand{\mstar}{$\rm M_{\star}$}

\newcommand{\fehgradrinf}{$\rm \nabla [Fe/H]_{R_{inf}}$}
\newcommand{\fehgradeinf}{$\rm \nabla [Fe/H]_{E_{inf}}$}
\newcommand{\fehgradrinfexp}{$\rm \nabla [Fe/H]_{R_{inf, exp}}$}
\newcommand{\fehgradrnow}{$\rm \nabla [Fe/H]_{R_{now}}$}
\newcommand{\fehgradenow}{$\rm \nabla [Fe/H]_{E_{now}}$}

\title[GES metallicity gradients]{From order to chaos: the blurred out metallicity gradient of the Gaia-Enceladus/Sausage progenitor}

\author[A. Carrillo et al.]{
Andreia Carrillo$^{1,2,3}$\thanks{E-mail: acarrillo@carleton.edu},
Alis J. Deason$^{2,3}$, Azadeh Fattahi$^{2,4}$, Robert J. J. Grand$^{5}$, and Francesca Fragkoudi$^{2}$ 
\\
$^{1}$Department of Physics and Astronomy, Carleton College, 1 North College St., Northfield, MN 55057, USA \\
$^{2}$Institute for Computational Cosmology, Department of Physics, Durham University, Durham DH1 3LE, UK \\
$^{3}$Centre for Extragalactic Astronomy, Department of Physics, University of Durham, South Road, Durham DH1 3LE, UK
\\
$^{4}$The Oskar Klein Centre, Department of Physics, Stockholm University, AlbaNova University Center, 106 91 Stockholm, Sweden
\\
$^{5}$Astrophysics Research Institute, Liverpool John Moores University, 146 Brownlow Hill, Liverpool, L3 5RF, UK\\
}

\date{Accepted XXX. Received YYY; in original form ZZZ}

\pubyear{2015}

\begin{document}
\label{firstpage}
\pagerange{\pageref{firstpage}--\pageref{lastpage}}
\maketitle

\begin{abstract}
The powerful combination of Gaia with other Milky Way large survey data has ushered in a deeper understanding of the assembly history of our Galaxy, which is marked by the accretion of Gaia-Enceladus/Sausage (GES). As a step towards reconstructing this significant merger, we examine the existence and destruction of its stellar metallicity gradient. We investigate 8 GES-like progenitors from the Auriga simulations and find that all have negative metallicity gradients at infall with a range of -0.09 to -0.03 dex/kpc against radius and -1.99 to -0.41 dex/$\rm 10^{-5} km^{2}s^{-2}$ against the stellar orbital energy. These gradients get blurred and become shallower when measured at $z=0$ in the Milky Way-like host. The percentage change in the radial metallicity gradient is consistently high (78-98\%), while the percentage change in the energy space varies much more (9-91\%). We also find that the most massive progenitors show the smallest changes in their energy metallicity gradients. At the same present-day galactocentric
radius, lower metallicity stars originate from the outskirts of the GES progenitor. Similarly, at fixed metallicity, stars at
higher galactocentric radii tend to originate from the GES outskirts. We find that the GES stellar mass, total mass, infall time,
and the present-day Milky Way total mass are correlated with the percentage change in metallicity gradient, both in radius and
in energy space. It is therefore vital to constrain these properties further to pin down the infall metallicity gradient of
the GES progenitor and understand the onset of such ordered chemistry at cosmic noon.



\end{abstract}




\section{Introduction}

The Milky Way is our most detailed laboratory for understanding the hierarchical formation of galaxies, owing to the wealth of photometric, astrometric, and spectroscopic data we can gather for its individually resolved stars. A deep image of the Milky Way halo reveals a field of stellar streams \citep{belokurov06,shipp18}, depicting current evidence of dwarf galaxies and globular clusters tidally disrupting due to the Milky Way's gravitational potential. 
Although these are systems that are disrupting \textit{today}, the Milky Way's voracious eating habit is true across time \citep{deason16}. From small samples (N$\lesssim$100) of local stars with chemistry and kinematics, many works \citep[e.g.][]{nissen00,fulbright02,nissen11,schuster12} found that there are chemo-kinematically distinct populations in the stellar halo that correspond to an in-situ and an accreted origin. Further out in the halo, \citet{deason18} found a break in the stellar halo density profile, which could be explained by a large dwarf merger whose stars' apocenters pile up at 20 kpc. Further investigation into this population with the advent of \textit{Gaia} data \citep{Gaia18} mapping $10^9$ stars in the Galaxy solidified that this was the Milky Way's last major merger that happened roughly 10 Gyr ago \citep{belokurov18,helmi18,haywood18}. This dwarf progenitor, which ultimately altered the course of our Galaxy's history \citep{belokurov20,bonaca20,grand20,ciuca22,dillamore22,buck23,merrow24}, has been dubbed the \textit{Gaia}-Enceladus/Sausage (GES).

It is undeniable how our own Galaxy's evolution is intimately entangled with the formation and eventual destruction of GES. Therefore, to learn more about the assembly history of the Milky Way, it is important to also understand the assembly of its most significant merger. Consequently, many works have reconstructed GES from the combination of large survey data and simulations. \citet{naidu21b} created a suite of N-body simulations and found that the best-fit GES progenitor to the H3 survey data \citep{conroy19} was accreted with a $15^{\circ}$ incline with respect to the Milky Way disk and a circularity of 0.5 (where 0 is perfectly radial). With this orbit, they also predict a double break in the stellar halo profile due to the two apocenter pile-ups of GES debris, which \citet{han22} later confirmed in detail with the same observations. Recent work by \citet{skuladottir25} agrees with this two-passage, double-break, scenario and the authors find that possible remnants from the outskirts of GES are less chemically evolved from detailed abundance ratio measurements.  
Multiple works have also reconstructed the star formation history of GES \citep[e.g.][]{vincenzo19,hasselquist21,ernandes24}, determining the particular chemical evolution and history the GES progenitor went through before it fully merged with the Milky Way.

An inevitable result of star formation and evolution is the chemical enrichment of the galaxy. Part of this history is eventually locked up in the stellar atmospheres, which helps our pursuit in chemically-tagging stars to their origin \citep{freeman02}. Indeed, many works have shown that the bulk of GES stars are chemically distinct in different elemental families compared to the Milky Way and its surviving satellite population \citep[e.g.][]{hayes18,carrillo22a, buder22}. However, we can probe even more deeply by investigating the metallicity gradient. The metallicity gradient serves as a window as to how the chemistry locked up in stars may be different at various galactic regions because of the inflow and outflow of gas, the environment, morphology, or star formation activity of the galaxy \citep{belfiore17,sanchez2020,mercado21}. Specifically for GES, it also aids in our knowledge of \textit{when} a satellite of its size could have started having this ordered chemistry, in contrast to much earlier snapshots in time when galaxies were clumpier, less ordered systems or proto-galaxies \citep{horta24}. This is becoming especially important as we find more lensed systems at higher redshifts (z $>$ 1) for which we can measure metallicity gradients to compare to \citep[e.g.][]{jones15,wang17,curti20}, albeit largely for gas instead of stars.

Many works have indeed tried to unravel the metallicity gradient of the GES progenitor. From their best-fit model and mapping the present-day angular momentum, $L_z$, to the mean pre-merger location of GES stars, \citet{naidu21b} reconstructed the stellar metallicity gradient of GES to be very weakly negative at $-0.016$ dex/kpc. These results are in congruence with  \citet{mori24}, who argue that the proposed distinct origin of more metal-poor retrograde substructures should be considered more carefully, as they may just be the metal-poor outskirts of the GES progenitor. \citet{khoperskov23} also reconstructed the infall metallicity gradient of GES using APOGEE DR17 \citep{apogeedr17} data in combination with N-body simulations and the HESTIA hydrodynamical cosmological simulations \citep{libeskind24}. They similarly found a (more) negative metallicity gradient of -0.1 dex/kpc that ultimately gets destroyed and becomes shallower at present-day by an order of magnitude. With next-generation surveys aiming to map the more distant halo such as DESI \citep{cooper23}, WEAVE \citep{jin24}, and 4MOST \citep{helmi194most}, we have access to more GES stars at larger galactocentric radii, the prime discovery space that allows for a more complete cartography of this satellite. From the high-redshift end, analogs of the Milky Way with a GES-like system nearby are predicted to be observable and within the same field of view with JWST at $z=2$ \citep{evans22}. It is therefore becoming more important and possible to unearth detailed information about the GES progenitor from both the local Universe and at higher redshifts. 

In this work, we use the Auriga hydrodynamical cosmological zoom simulations of Milky Way-like halos \citep{grand17} to 
explore the metallicity distributions of GES-like systems before and after they are engulfed by the Milky Way. We use a subsample of halos whose stars resemble the kinematics of GES stars in the observations \citep{fattahi19}. Our analysis is guided by the following key questions: (1) Would an ancient satellite like GES already have ordered chemistry before it was accreted? (2) How does an initial metallicity gradient get destroyed? (3) Can we unwind the dynamical evolution of the GES stars and retrieve the progenitor's metallicity gradient? We attempt to answer these questions and organize this paper as follows: Section \ref{sec:simulation_data} desribes the simulation data, Section \ref{sec:pre_vs_post} compares pre-merger and post-merger metallicity gradients of stars belonging to GES, Section \ref{sec:metgrad_change} outlines which merger properties affect the destruction of the infall metallicity gradient, Section \ref{sec:discussion} shows the implications of our results in the context of observational data of similar mass galaxies, and Section \ref{sec:conclusion} summarizes our main takeaways.




\section{Simulation Data}
\label{sec:simulation_data}

\begin{table*}
\begin{center}
\caption{Auriga halos with GES-like systems as identified in \citet{fattahi19}. The column titles are as follows: (1) Auriga halo with a GES progenitor (2)~GES progenitor \mstar~(infall, peak) (3) GES progenitor total mass $\rm M_{total}$ (i.e., $\rm M_{200}$; infall, peak) (4) Host's peak $\rm M_{\star}$ (5) Host's peak $\rm M_{total}$ (6) Infall time (from Big Bang) of GES (7) GES Disc-to-total mass ratio as calculated in \citet{orkney23}. In our analysis in Section \ref{sec:metgrad_change}, we use the infall \mstar~and $\rm M_{total}$.} 
\begin{tabular}{ccccccc}
\hline \hline
(1) & (2) & (3) & (4) & (5) & (6) & (7) \\
Halo & GES $\rm M_{\star}$ & GES $\rm M_{total}$ & Host $\rm M_{\star}$ & Host $\rm M_{total}$ & Infall time & $D/T$   \\

 & $10^{9}\rm M_{\odot}$ & $10^{11}\rm M_{\odot}$ & $10^{11}\rm M_{\odot}$ & $10^{12}\rm M_{\odot}$ & Gyr &  \\
\hline
Au-5 & 3.42, 3.83 & 0.76, 1.26 & 0.71 & 1.19 & 6.20 & 0.10  \\
Au-9 & 1.76, 1.88 & 0.70, 1.76 & 0.63 & 1.16 & 3.45 & 0.30   \\
Au-10 & 0.92, 0.97 & 0.34, 0.39 & 0.62 & 1.02 & 5.92 & 0.77   \\
Au-15 & 2.27, 2.53 & 1.02, 1.26 & 0.43 & 1.04 & 6.36 & 0.57  \\
Au-17 & 0.37, 0.38 & 0.22, 0.33 & 0.79 & 1.02 & 2.68 & 0.20  \\
Au-18 & 1.39, 1.44 & 0.38, 0.75 & 0.84 & 1.39 & 4.51 & 0.10   \\
Au-24 & 2.40, 2.56 & 0.76, 1.09 & 0.77 & 1.57 & 4.99 & 0.48   \\
Au-27 & 4.08, 4.08 & 0.77, 1.72 & 1.03 & 1.85 & 4.21 & 0.57   \\
\hline
\hline
\end{tabular}
\label{tab:auriga_halos}
\end{center}
\end{table*}

We use the Auriga hydrodynamical simulations \citep{grand17}, that include 30 high-resolution, cosmological zoom-in simulations of Milky Way-mass halos. These halos were selected to have a virial mass of 
$\rm 1-2 \times 10^{12}~M_{\odot}$ from the dark matter-only $\rm 100^{3}~Mpc^{3}$ periodic box of the EAGLE project \citep{schaye15,crain15} and resimulated with the \textsc{AREPO} code \citep{springel10}. The cosmological parameters were adopted from the Planck Collaboration \citep{planck14}. This work uses the Auriga Level 4 runs with mass resolution of $\sim 3 \times 10^{5} \rm M_{\odot}$ for the dark matter particles and $\sim 5 \times 10^{4} \rm M_{\odot}$ for the gas cells.

Auriga incorporates a comprehensive model for galaxy formation physics (refer to \citealt{grand17} for more details). This includes primordial and metal cooling, star formation and stellar feedback, and chemical enrichment from core-collapse supernovae, Type Ia supernovae, and winds from asymptotic giant branch stars. Auriga also includes a sub-grid model for the interstellar medium, processes for black hole formation and feedback, a uniform photoionizing UV background, and the effects of magnetic fields. 

We specifically use this simulation suite as \citet{fattahi19} identified halos in Auriga that contain a GES-like merger. Here we briefly describe this work but for more details, we refer to \citet{fattahi19} and to \citet{orkney23} who explored the diverse properties of Auriga halos with such GES progenitors. 

In \citet{fattahi19}, star particles were classified as "in-situ" if they were bound to the main progenitor of the Milky Way analogue at their formation time, as determined by the \textsc{SUBFIND} algorithm \citep{springel01}. This association is made at the snapshot immediately following their birth. If the formation occurred at $z>3$, the $z=3$ association is used. Star particles that are bound to the main halo at $z=0$ but were originally formed in a different halo 
were classified as "accreted" \footnote{From this definition, stars that formed from the gas stripped from the satellite are considered in-situ.}.  

Within the Auriga simulations, \citet{fattahi19} identified a subset of halos containing accreted stars characterized by high orbital anisotropy ($\beta > 0.8$) and high metallicity ([Fe/H] $\sim$ -1), similar to the GES observed in the Milky Way. The accreted stars contributing to the highly-anisotropic ``sausage" feature in the \Vphi - \Vr~in the simulations is typically dominated by a single progenitor. In most cases, this progenitor is the most massive contributor to the halo, with a stellar mass of 
$10^9-10^{10}~\rm M_{\odot}$  and accreted 6-10 billion years ago. We use this subset of Auriga halos to explore the metallicity gradient evolution of GES. The properties of these halos are detailed in Table \ref{tab:auriga_halos}. 
We note that Au-10 actually contains two GES-like mergers that happened at the same time (see \citealt{orkney23} for details) but we consider only one of these progenitors in our analysis. With this sample of GES-like progenitors in Auriga and their snapshots through time, we now investigate the distribution of the associated stars and their metallicities. 



\section{Pre-merger vs Post-merger of GES}
\label{sec:pre_vs_post}


We use the $z=0$ snapshot to represent the post-merger GES remnant in the simulations in order to be consistent with the observations. However, defining the pre-merger GES is more complex. To ensure consistency across different halos, we consider the pre-merger GES at infall time (i.e., when the progenitor crosses the virial radius of the host), as indicated in Table \ref{tab:auriga_halos}. 

\begin{figure*}
\includegraphics[width=\textwidth]{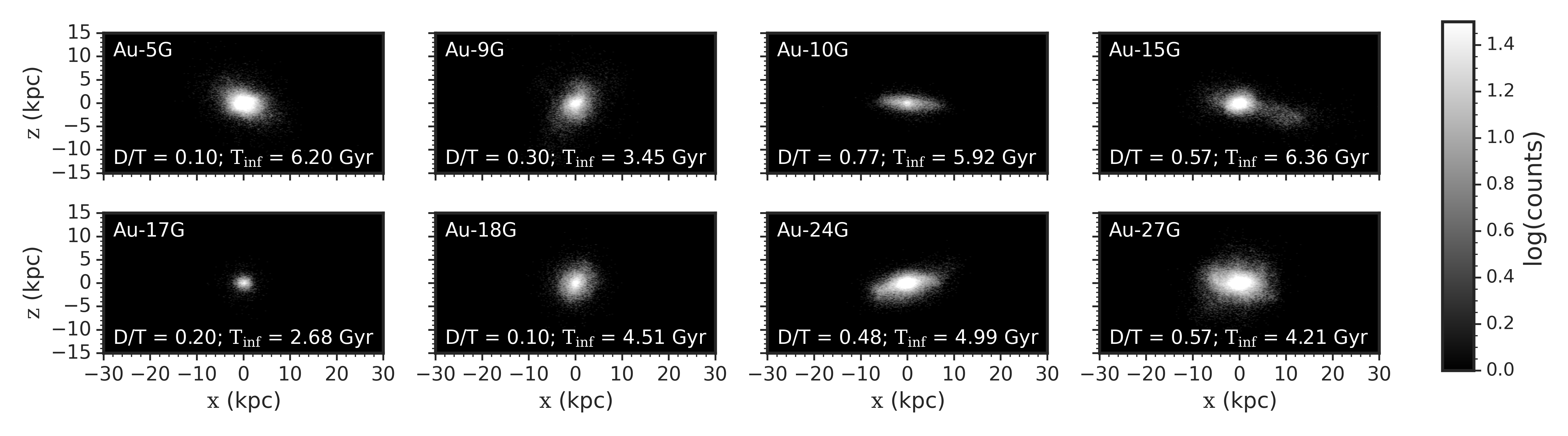}
 \caption{\textbf{Density plots of Auriga GES progenitors in the XZ projection}. Each subplot is labeled at the top with the GES-like halo in Auriga as well as the disc-to-total mass ratio, $D/T$, from \citet{orkney23} and the infall time (from Big Bang) at the bottom. There is a diversity in the shapes of these progenitors: some have rounder morphology while others are flatter and more disk-like.}
 \label{fig:ges_xz}
\end{figure*}

With the GES main progenitor established, we then translate and rotate the orientation of the simulation box using the angular momentum of these GES stars at infall, such that both positions and velocities are centered on the GES progenitor. These GES progenitors are shown in Figure \ref{fig:ges_xz} in the X-Z orientation and clearly exhibit a diversity in size and morphology as thoroughly discussed in \citet{orkney23}. The authors have also calculated the disc-to-total mass ratio ($D/T$) for these systems based on the probability of a star particle belonging to the disc or halo given its kinematics. We include this important physical property in Table \ref{tab:auriga_halos} as well as labels in Figure \ref{fig:ges_xz} to emphasize the range of ``disky-ness'' the GES progenitor may have had. 
As shorthand, we denote the GES progenitor in each Auriga halo with a "G". To highlight a few examples, Au-5G shows more pressure support both in its morphology and $D/T$. Au-10G, on the other hand, has a more flattened morphology in the z direction which is corroborated by the high $D/T$. Au-15G shows tidal disruption as it fell in with another subhalo that it later merges with, and both Au-24G and Au-27G exhibit similar signatures but to a lesser degree. We note, however, that these disrupting features are more easily identified in the morphology of progenitors with higher $D/T$ (i.e., $D/T \gtrsim 0.50$).

Defining the center of the GES-like system through time 
needs to be considered carefully to ensure that the pre- and post-merger comparisons are as direct as can be. We therefore identify the most bound star particles\footnote{This number varies from halo to halo but it was ensured that $N > 1000$.} at infall and use their center of mass as the center of the GES at all snapshots, from infall to present-day. In practice, this coincides with the center of the GES progenitor at infall for the pre-merger scenario and the center of the host at $z=0$ in the case of the post-merger.   

\subsection{Pre-merger GES properties}
\label{sec:premerger}

\begin{figure*}
\includegraphics[width=\textwidth]{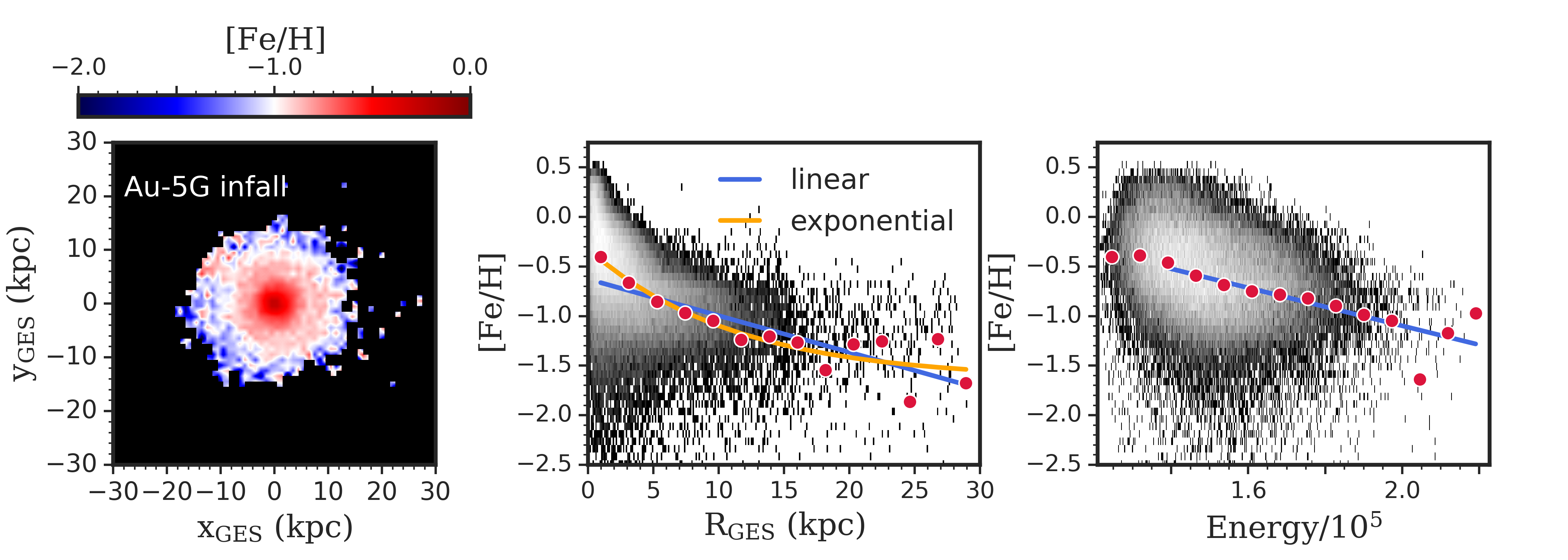}
\includegraphics[width=\textwidth]{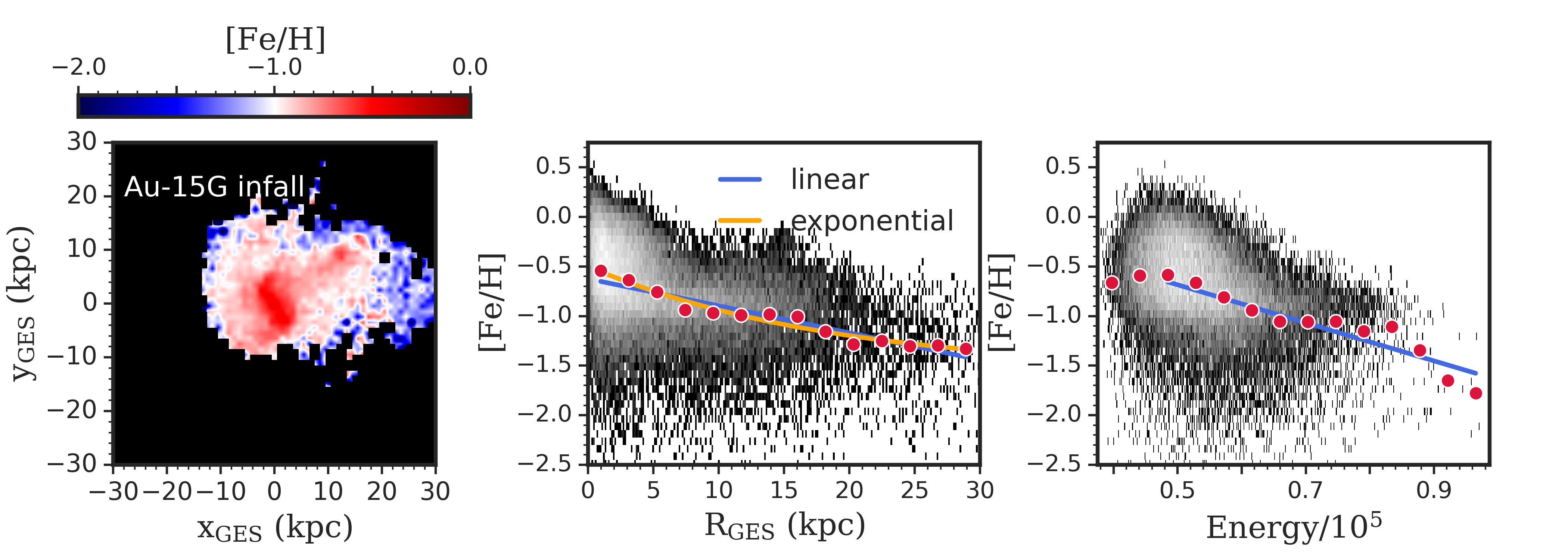}
 \caption{\textbf{Pre-merger picture of GES in Auriga.}  The sub-panels contain the XY projection of the GES progenitor colored by the average [Fe/H] along the z axis in the first column where redder means higher [Fe/H], the 2D histogram of [Fe/H] vs $\rm R_{GES}$ in the second column, and the 2D histogram of [Fe/H] vs orbital energy in the third column. The mean values of [Fe/H] are denoted with red circles for the trends against  $\rm R_{GES}$ and orbital energy, with the best-fit linear (blue) and exponential (orange) relations over-plotted. In both examples, Au-5G (top row) and Au-15G (bottom row), the GES progenitor shows the existence of ordered star formation, with higher [Fe/H] in the most bound inner regions which then decreases towards the less bound outskirts. The scatter in [Fe/H] also decreases with increasing R.}
 \label{fig:premerge}
\end{figure*}

We take the discussion of the pre-merger picture of GES further by going beyond its morphology and looking at its chemical properties. Specifically, we explore how the metallicity, [Fe/H], of a star particle might trace different spatial and dynamical regions of the GES progenitor before infall, giving us a potential tool to reconstruct this significant merger.  

The [Fe/H] is tracked in the simulations and scaled to Solar values following \citet{asplund09}. The [Fe/H] from the simulations are noticeably higher compared to the GES in the real Milky Way, and this is a known discrepancy \citep[e.g.][]{fattahi19, grand21, orkney23,carrillo24}. However, this has a minimal effect on this work as we are assessing relative trends, i.e. the [Fe/H] gradient with respect to radius or orbital energy, instead of absolute values. We define the GES-centric radius, $\rm R_{GES}$, as the cylindrical radius of star particles post-translation and rotation. In addition, we calculate the orbital energy\footnote{When we say energy in this paper, we also mean the orbital energy i.e., the sum of the kinetic and potential energies.} of each star particle by summing its kinetic and potential energy as given in the simulations.  We derive the kinetic energy from the re-oriented velocities centered on GES but note that the potential is taken directly from the simulation box. Therefore, the resulting value for the orbital energy may not necessarily be negative, as one would expect for stars that are still bound to the GES progenitor. For a more detailed discussion, see \citet{grand24}. In practice, the correction at a given snapshot is a constant value of energy added to (or subtracted from) all particles, hence the slope is not affected.
We note that our calculation for the orbital energy for both the pre-merger and post-merger scenario is the same, allowing us to compare the two. 

We now 
examine the pre-merger metallicity distribution of the main progenitor of GES in Auriga as shown in Figure \ref{fig:premerge}. Here we highlight Au-5G and Au-15G. In the first column, the XY projections colored by [Fe/H] visually shows how the distribution of [Fe/H] traces that of the stellar material, and therefore also any morphological disturbances they encounter. Specifically, Au-15G falls in with a companion, which causes some disruption to the GES progenitor as is evident in its morphology.  

The [Fe/H] vs $\rm R_{GES}$ trends in the second column further support this narrative. The undisturbed Au-5G has a more steeply declining [Fe/H] with $\rm R_{GES}$ that becomes sparse and flattens out at 11 kpc. The Au-15G system, on the other hand, is already perturbed at infall by a companion\footnote{In a case like this, the most obviously massive and therefore dominant progenitor
linked with GES is chosen as the pre-merger system.}, showing a steep decline in [Fe/H] with $\rm R_{GES}$ up to 6 kpc, but then stays constant and stretches out to 16 kpc, until finally falling off again beyond this distance. It is interesting to highlight this system because it gives a glimpse of what will happen to the metallicity distribution of the GES systems in the simulations when they get tidally disrupted by the host i.e., a stretching and washing away of the original metallicity gradient. In addition, for both progenitors showcased (and in fact for the rest of the sample), the scatter in [Fe/H] is largest in the center and smallest in the outskirts, indicative of the different star formation histories in these regions.

Lastly, the [Fe/H] gradients with orbital energy for Au-5G and Au-15G are also both negative, with lower [Fe/H] for less bound star particles. The [Fe/H] trend with energy for Au-15G shows signatures of disruption (or at least non-uniformity), similar to the gradient with $\rm R_{GES}$. 
We fit a linear relationship to these GES infall [Fe/H] gradients with $\rm R_{GES}$ and energy, denoted as \fehgradrinf~and~\fehgradeinf, respectively. We also fit an exponential function to the [Fe/H] gradient with $\rm R_{GES}$ of the form $\rm [Fe/H] $$= Ae^{B \rm \times R_{GES}} + C$, and report the exponent, $B$, as \fehgradrinfexp. For now, we list these values in Table \ref{tab:gradients} but will further discuss them in Section \ref{sec:metgrad_change} especially in the context of the change in metallicity gradient with time. 

\begin{table*}
\begin{center}
\caption{Metallicity gradient of GES-like halos in Auriga at infall and present day. The columns are ordered as follows: (1) Auriga halo with a GES progenitor (2) slope of the linear fit to the infall [Fe/H] gradient with respect to $\rm R_{GES}$, \fehgradrinf~3) exponent of the exponential fit to the infall [Fe/H] gradient with respect to $\rm R_{GES}$, \fehgradrinfexp~(4) slope of the linear fit to the infall [Fe/H] gradient with respect to orbital energy, \fehgradeinf~(5)  slope of the linear fit to the present-day [Fe/H] gradient with respect to $\rm R_{host}$, \fehgradrnow~(6) slope of the linear fit to the present-day [Fe/H] gradient with respect to orbital energy, \fehgradenow~(7) percentage change in the radial [Fe/H] gradient from infall to today (8) percentage change in the energy [Fe/H] gradient from infall to today.  } 
\begin{tabular}{cccccccc}
\hline \hline
(1) & (2) & (3) & (4) & (5) & (6) & (7) & (8)  \\
halo & \fehgradrinf & \fehgradrinfexp & \fehgradeinf & \fehgradrnow & \fehgradenow & $\rm \Delta_{\nabla[Fe/H]_R}$ & $\rm \Delta_{\nabla[Fe/H]_E}$ \\

 & dex/kpc & exponent, $\rm kpc^{-1}$ & dex/$\rm 10^{-5} km^{2} s^{-2}$ & dex/kpc & dex/$\rm 10^{-5} km^{2} s^{-2}$ & \% & \% \\
\hline
Au-5G & -0.0368 $\pm$ 0.0062 & -0.0891 $\pm$ 0.0429 & -0.9587 $\pm$ 0.1574 & -0.0066 $\pm$ 0.0007 & -0.6551 $\pm$ 0.0690 & 82 & 32 \\
Au-9G & -0.0284 $\pm$ -0.0043 & -0.0983 $\pm$ 0.0316 & -0.5875 $\pm$ 0.0744 & -0.0062 $\pm$ 0.0007 & -0.4224 $\pm$ 0.0235 & 78 & 28 \\
Au-10G & -0.0563 $\pm$ 0.0046 & - & -1.9915 $\pm$ 0.3824 & -0.0014 $\pm$ 0.0003 & -0.1872 $\pm$ 0.0381 & 98 & 91 \\
Au-15G & -0.0272 $\pm$ 0.0023 & -0.0532 $\pm$ 0.0172 & -1.9183 $\pm$ 0.1873 & -0.0060 $\pm$ 0.0013 & -0.7149 $\pm$ 0.0476 & 78 & 63 \\
Au-17G & -0.0925 $\pm$ 0.0139 & - & -0.8672 $\pm$ 0.1998 & -0.0046 $\pm$ 0.0007 & -0.5647 $\pm$ 0.1081 & 95 & 35 \\
Au-18G & -0.0448 $\pm$ 0.0291 & -0.1375 $\pm$ 0.2215 & -0.9567 $\pm$ 0.2969 & -0.0052 $\pm$ 0.0005 & -0.4826 $\pm$ 0.0685 & 88 & 50 \\
Au-24G & -0.0392 $\pm$ 0.0049 & -0.0585 $\pm$ 0.0288 & -0.5879 $\pm$ 0.1130 & -0.0038 $\pm$ 0.0006 & -0.4944 $\pm$ 0.0266 & 90 & 16 \\
Au-27G & -0.0300 $\pm$ 0.0044 & -0.1093 $\pm$ 0.0274 & -0.4081 $\pm$ 0.1168 & -0.0066 $\pm$ 0.0007 & -0.3734 $\pm$ 0.0128 & 78 & 9 \\

\hline
\hline
\end{tabular}
\label{tab:gradients}
\end{center}
\end{table*}

From this exploration, regardless of the details of the halo---when they fell in, their stellar and total mass, even the prominence of the disc---it is undeniable that these GES-like progenitors already had some semblance of ordered star formation before they fell into their host galaxy, as reflected by their stellar metallicity and their negative slopes against radius and energy.

\subsection{Post-merger GES properties}
\label{sec:postmerger}


\begin{figure*}
\includegraphics[width=\textwidth]{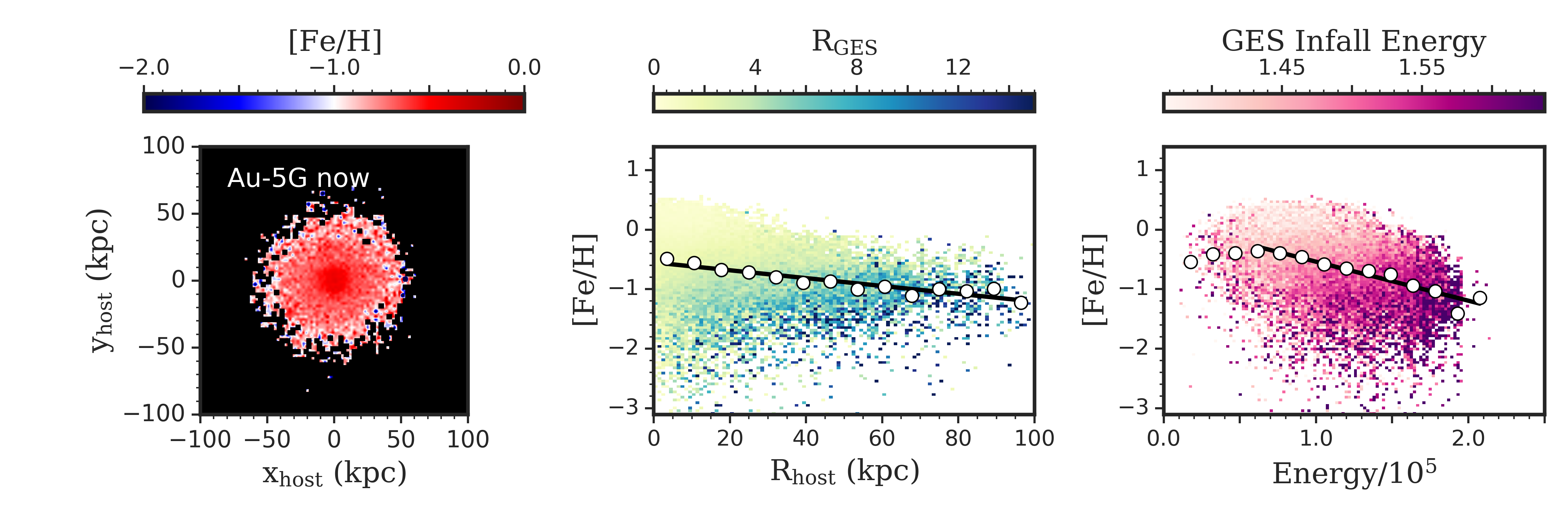}
\includegraphics[width=\textwidth]{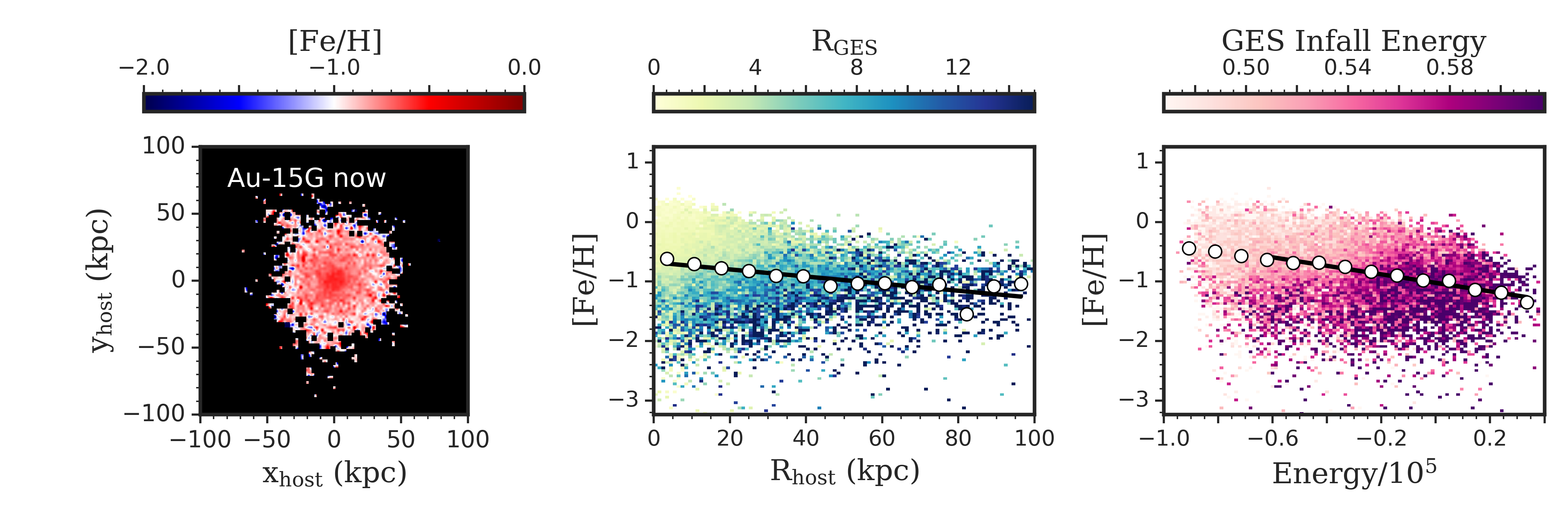}
 \caption{\textbf{Post-merger picture of the GES debris in the host galaxy.} This figure is similarly arranged as Figure \ref{fig:premerge} and shows the XY projection of the GES debris at $z=0$ colored by [Fe/H] in the first column, the [Fe/H] vs $\rm R_{host}$ in the second column, and the [Fe/H] vs orbital energy in the third column for Au-5G (top row) and Au-15G (bottom row), all with respect to the host galaxy at $z=0$. The median values of [Fe/H] as a function of $\rm R_{host}$ and energy are marked with circles with the best-fit linear trend over-plotted (black solid line). The [Fe/H] vs $\rm R_{host}$ panel is colored by $\rm R_{GES}$ while the [Fe/H] vs present-day orbital energy panel is colored by the orbital energy in the GES progenitor. Although the originally steeper [Fe/H] gradient is washed away, some signature of it still remains as the [Fe/H] still declines with higher $\rm R_{host}$ and energy. Additionally, at a given $\rm R_{host}$, stars with lower [Fe/H] tend to have come from larger $\rm R_{GES}$ at infall.} 
 \label{fig:postmerge}
\end{figure*}


We now fast-forward to the present day and show the $z=0$ snapshot of the same halos, Au-5G and Au-15G, in Figure \ref{fig:postmerge}. The sub-panels are ordered analogously to those of Figure \ref{fig:premerge} but now centered on the Milky Way-like hosts in the simulations.  In addition, the panels that show [Fe/H] vs galacto-centric distance, $\rm R_{host}$, and present day orbital energy, are colored by $\rm R_{GES}$ and orbital energy in the GES progenitor at infall, respectively. We note that massive satellites like the GES could continue forming stars after infall. To ensure we are comparing the same population, we are only considering the stars that are cross-matched between infall and present-day. 

Starting with the XY projection (first column), we can see that the high-metallicity stars (i.e., [Fe/H] > -1) of the GES debris dominate at any given region. This is largely due to these stars being more prevalent as shown in the 2D histograms in Figure \ref{fig:premerge}.    
However, there is still a slight gradient, with the more centrally-located GES stars having higher metallicity compared to those farther out in the halo. This present-day gradient appears to be shallower compared to the GES-centric gradient in Figure \ref{fig:premerge} which more visibly shows that the outskirts have lower [Fe/H]. 

Nonetheless, a gradient still persists, which is more clearly seen in the second column showing [Fe/H] vs $\rm R_{host}$. The mean [Fe/H] in different $\rm R_{host}$ bins are shown with circles, highlighting the decrease in [Fe/H] with increasing $\rm R_{host}$.  We also fit a linear relationship to [Fe/H] vs $\rm R_{host}$, and include this obtained gradient, \fehgradrnow, in Table \ref{tab:gradients}. It is worth noting that although the original [Fe/H] gradient has been washed away, the negative sign of the slope is retained. The stars may be phase-mixed but which stars end up where is not random; those that formed at smaller $\rm R_{GES}$ tend to be deposited at smaller $\rm R_{host}$. In addition, at the same $\rm R_{host}$, the stars with lower [Fe/H] tend to have come from larger distances within the progenitor at infall. Both these trends seem to hold true across all the halos included in our sample, albeit to varying degrees. 

Lastly, we investigate the [Fe/H] vs orbital energy at present day as shown in the last column of Figure \ref{fig:postmerge}. We similarly derive the orbital energy by summing the kinetic energy of stars centered and oriented with respect to the host galaxy, and the potential taken directly from the simulation box. Although the resulting orbital energy is largely negative, as the GES stars are now bound to the main halo, this is not always the case as seen for Au-5G in Figure \ref{fig:postmerge}. Nonetheless, as discussed in the previous section, this should not affect our calculation of the energy metallicity gradient.
This is also colored by the 16th to 84th percentile of the orbital energy in the GES progenitor at infall and the mean [Fe/H] in different energy bins shown with circle symbols. Similar to the second column, there is a negative trend in [Fe/H] with energy, wherein stars that are less bound to the host have lower [Fe/H]. This pattern is consistent throughout the rest of our sample and we quantify its slope, \fehgradenow, as listed in Table \ref{tab:gradients}. Both Au-5G and Au-15G indicate that stars that were less bound to GES prior to the merger remain less bound to the Milky Way-mass host in Auriga. However, this relationship is not consistently observed in other halos. For example, Au-10G and Au-17G exhibit a more mixed distribution of GES infall orbital energies across present-day orbital energies (see Appendix \ref{app:other_halos}). 

This exploration of the present-day [Fe/H] gradient with $\rm R_{host}$ and energy have shown that the pre-merger properties of GES could possibly be teased out from the data. The combination of  galactocentric radius and [Fe/H] at present-day is particularly promising in deriving GES-centric radius at infall. On the other hand, the link between present-day and infall orbital energies is not as straightforward. In the next section, we explore stars with different orbital energies in a sample GES progenitor in more detail, both in terms of their chemistry and how they get stripped over time. 

\subsection{The evolution of $E-L_z$ and [Fe/H] over time}
\label{sec:elz}

\begin{figure*}
\includegraphics[width=\textwidth]{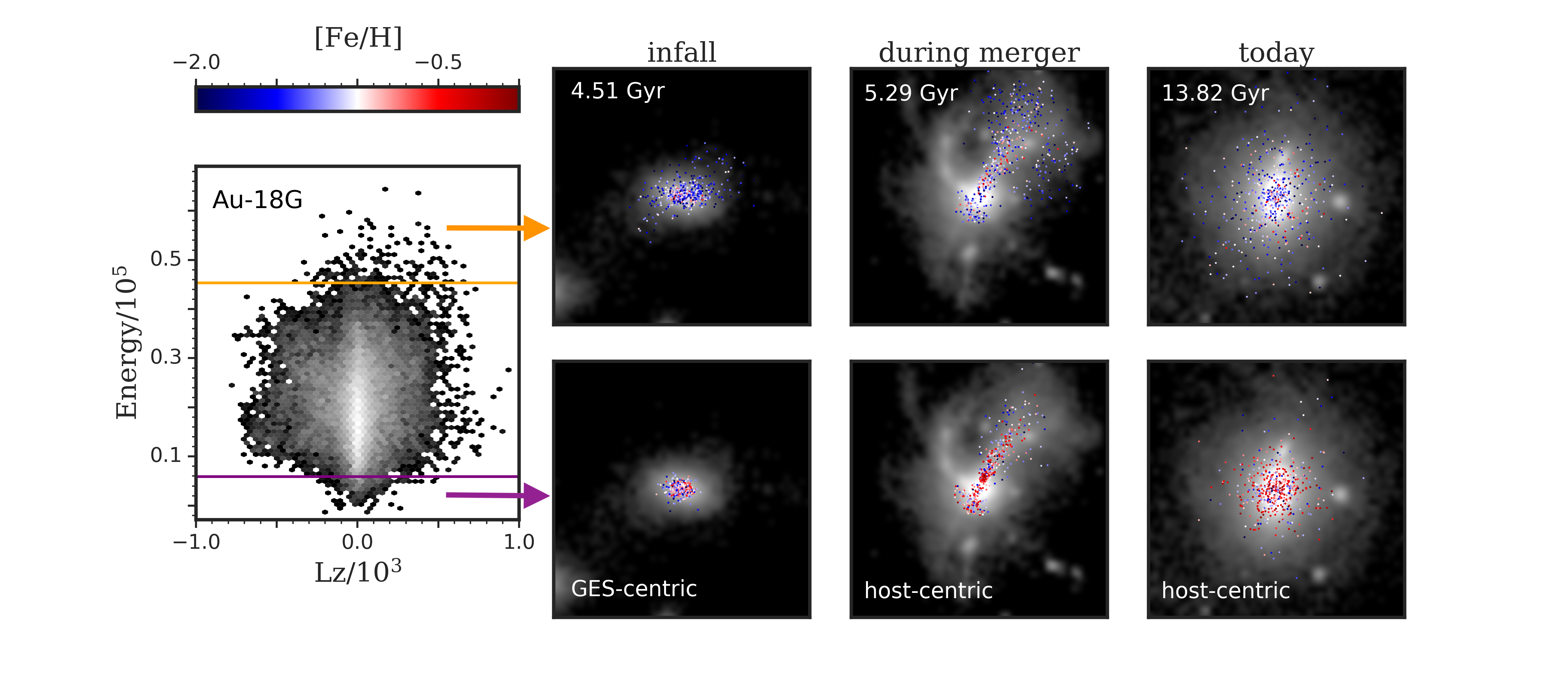}
 \caption{ \textbf{The evolution of the spatial distribution of GES stars with different infall orbital energies.} We show Au-18G as an example but emphasize that the rest of the GES progenitors in Auriga exhibit similar trends. First column shows the $E-L_z$ in the GES progenitor at infall with the $\rm 2^{nd}$ (purple line) and $\rm 98^{th}$ (orange line) percentiles marked. We use these demarcations to create our most bound and least bound samples, respectively, which we trace through time. Columns 2-4 show the evolution of the spatial distribution of stars in these two samples, where the top row shows the least bound stars and the bottom row shows the most bound stars. Column 2 portrays these different samples at infall centered on the GES main progenitor (60 kpc on each side), Column 3 during the merger (200 kpc a side), and Column 4 as we see the debris today (200 kpc a side). The star particles are coloured by their [Fe/H] values. Note that we center on the GES progenitor at infall, but center on the Milky Way-like host during the merger and at $z=0$ for ease of comparison.The surface density in gray are all scaled the same for Columns 2-4.} 
 \label{fig:elz}
\end{figure*}

We now investigate the properties of stars with different orbital energies, $E$, in the GES progenitor, and follow the evolution of their spatial distribution from infall all the way to $z=0$.  We highlight this evolution for Au-18G in Figure \ref{fig:elz} as an example but note that the other seven halos show similar trends. The left-most panel shows the $E-L_z$ diagram for the GES progenitor from which we selected the samples of most and least bound stars. Using the distribution of orbital energy of GES stars before infall, we consider the most bound stars to be those that fall under the $\rm 2^{nd}$ percentile i.e., anything below the purple line. The least bound stars are selected to be above the $\rm 98^{th}$ percentile i.e., anything above the orange line. Although the orbital energies are positive, the relative trend of which stars are the most bound vs least bound is preserved. In the next three columns, we follow the evolution of the spatial distribution of these same stars at different snapshots in time: at infall (second column), during the merger (third column), and phase-mixed today (fourth column), all colored by the star particle's [Fe/H], and separated into the least bound sample (top row) and most bound sample (bottom row). The subpanels on the spatial distributions of stars show surface densities in gray (all scaled the same) and encompass 60 kpc a side for the second column and 200 kpc a side for the third and fourth columns. We note that the 2nd-4th columns are mostly for illustrative purposes; therefore, for ease of comparison, we center on the GES progenitor for the infall picture, but center on the Milky Way-mass host during the merger and today. 

At infall, the most-bound stars are more compact and spherical in their spatial distribution compared to the least bound stars and they also have higher metallicities. 
The more spatially extended star particles for the least bound sample are more metal-poor and appear to be experiencing tidal disruption. 

During the merger, 0.8 Gyr after infall, we see a larger difference in how the most and least bound stars get stripped from the progenitor. The stars from the least bound sample are distributed more widely in the galaxy while the most bound stars have a narrower distribution that more closely follows the track of the progenitor. The differences in the [Fe/H] also become more apparent. For the least bound sample, the stars that are farther out in the galaxy are the ones that are more metal-poor while the ones along the GES track are slightly metal-richer. This is similar to what could be seen for the most bound sample. In fact, the core of the GES progenitor is visibly compact and more-metal rich, and the arms emanating from it have stars that are slightly metal-poorer than the core.

Lastly, we view the GES debris in the present-day, more than 9 Gyr after infall. Both the most bound and least bound material is now fully phase-mixed and seen at all radii. However, there are still some subtle differences between how the star particles settle into the host galaxy. For both samples, the majority of the stars are deposited within the inner regions ($\lesssim 20$ kpc), largely owing to dynamical friction dragging the massive progenitor towards the centre of the host \citep{Chandrasekhar1943}. The least bound sample have a higher fraction of stars in the outer halo of the galaxy than the most bound sample, because these less bound stars are stripped earlier in the disruption of the GES. Specifically, 50\% of the stars in the most bound population are in the inner 20 kpc while those in the least bound population are contained within 32 kpc. The final column of Figure \ref{fig:elz} also shows that the less bound GES stars have lower [Fe/H] overall compared to the most bound stars, especially if we compare them at the same distance from the host.

This investigation of the evolution of stars with different orbital energies in the GES progenitor has shown that there are larger differences at infall and during the merger, but these differences become less distinct as the debris gets phase-mixed. However, it is promising to see that indeed, at the same location in the host galaxy, a GES star with lower [Fe/H] most likely had higher orbital energy in the original progenitor. In addition, and perhaps unsurprisingly, the GES stars that are found farther out in the halo are lower in [Fe/H] and were less bound in the GES progenitor. 
\\
\\
In this section, we compared the pre-merger and post-merger picture of the GES progenitors in Auriga by investigating their morphology and how their stellar metallicities are distributed in location and orbital energy. From our qualitative comparisons, it seems encouraging that not all pre-merger properties are lost, as they vaguely track the similar present-day properties, specifically $\rm R_{host}$ and energy. In the next section, we further examine this and quantify the change in [Fe/H] gradient between the pre-merger and post-merger snapshots, and investigate if the degree to which the gradient changes could be due to other fundamental properties of the merger event. 


\section{Changes in the GES metallicity gradient over time}
\label{sec:metgrad_change}

We now examine the evolution of the simulated GES systems by investigating the change in metallicity gradient for all the sample. We first explore the relationship and differences between the pre-merger vs post-merger radial and energy [Fe/H] gradients in the simulations, and then interpret these trends in the context of the GES merger properties.

\subsection{Pre-merger vs post-merger [Fe/H] gradient}
\label{sec:pre_vs_post_grad}

\begin{figure}
\includegraphics[width=0.48\textwidth]{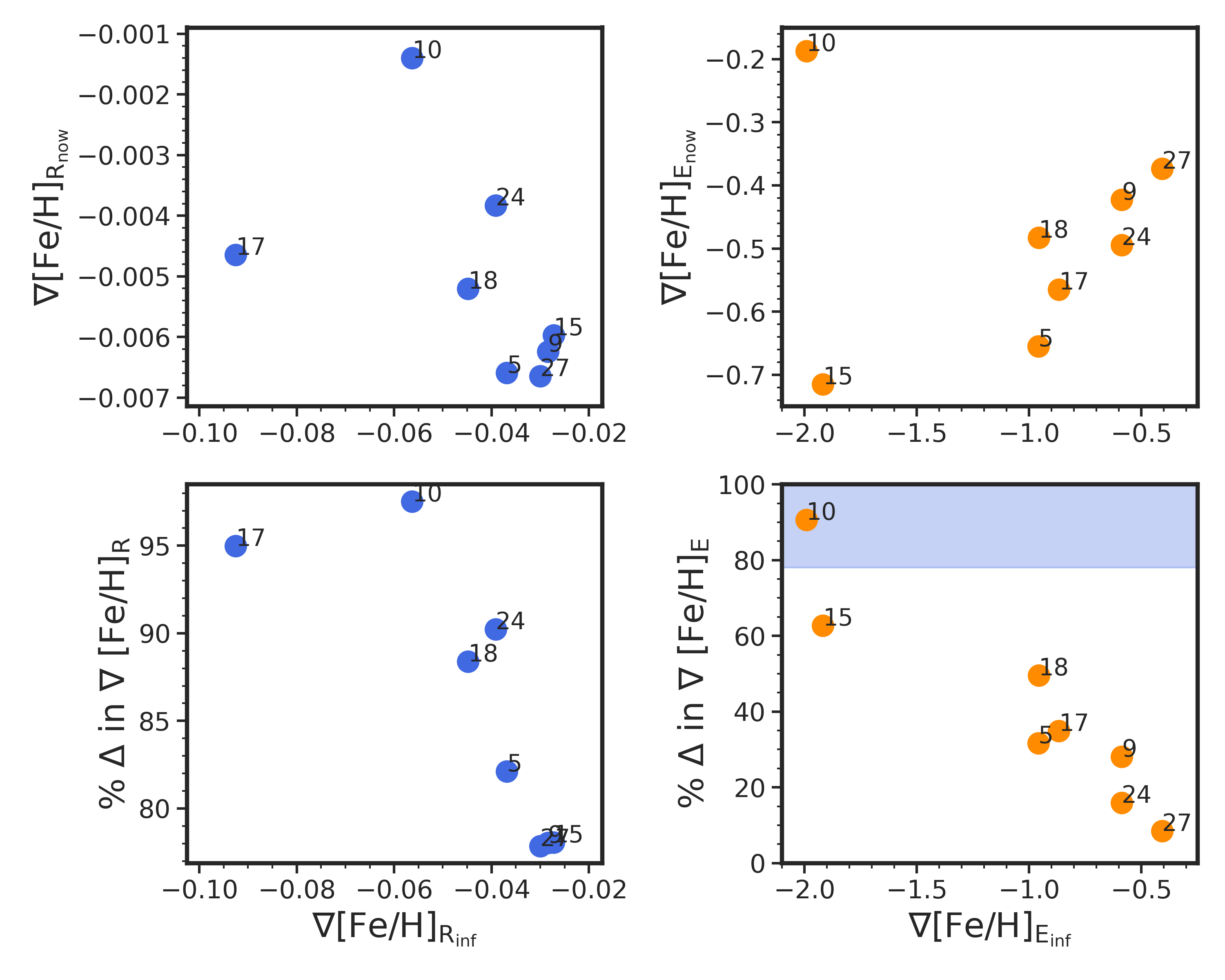}
\includegraphics[width=0.48\textwidth]{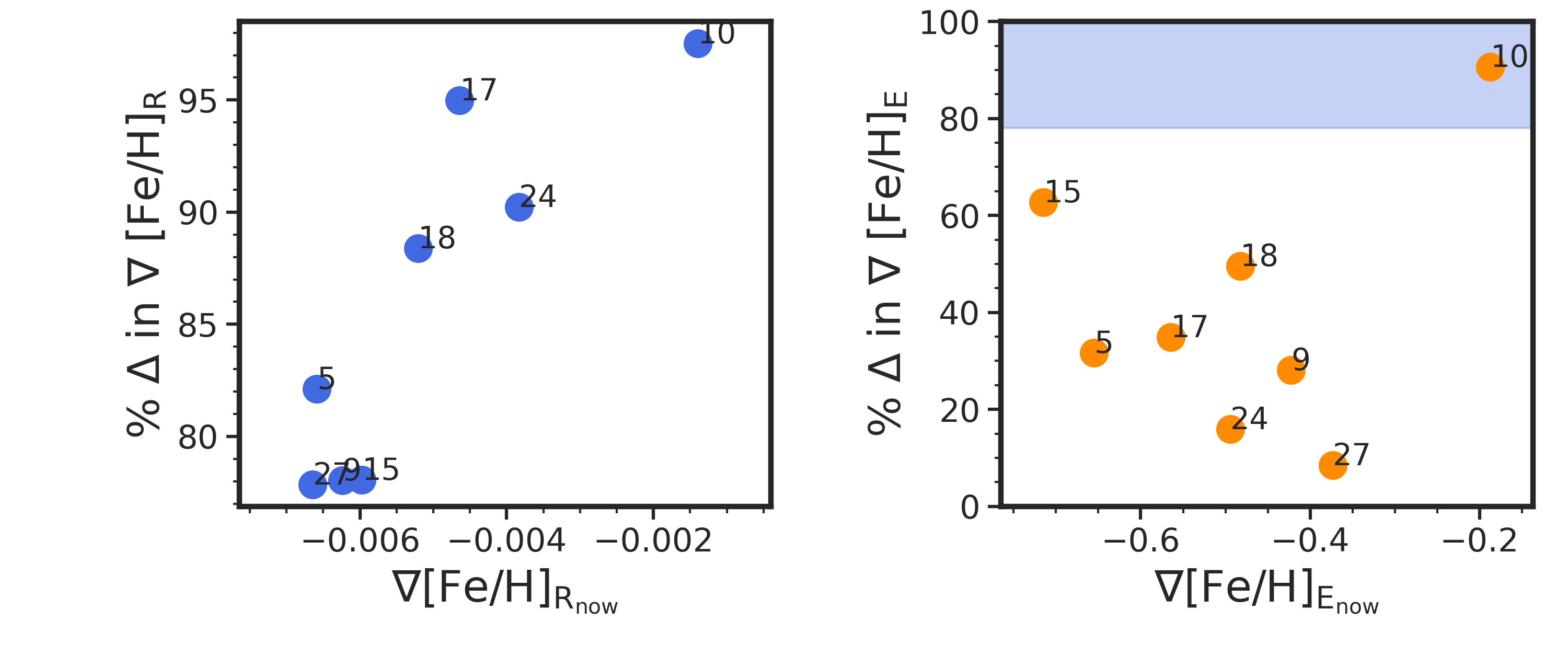}
 \caption{ \textbf{GES metallicity gradients in Auriga before and after merging.} The values plotted here are the slopes to the best linear fit. For the top two rows, the x axes show the infall [Fe/H] gradient with respect to $\rm R_{GES}$ (left) and orbital energy (right), while the bottom row shows these gradients as measured at present-day. The y axes in the top panels show the present-day radial (left) and energy (right) [Fe/H] gradients, and those in the middle and bottom panels show the percentage change in the radial (left) and energy (right) [Fe/H] gradients between infall and today. The blue-shaded regions indicate the range of the percentage change for the radial [Fe/H] gradient. The progenitors with steeper initial radial gradients tend to be shallower today, while those with steeper initial energy gradients typically remain steeper. In both radial and energy space, GES progenitors that had shallower initial gradients experience the least change. Our present-day measurement of the gradients are a potential window to knowing how much they have changed over time. }
 \label{fig:postvpre_grad}
\end{figure}

Figure \ref{fig:postvpre_grad} highlights the changes in the radial and energy [Fe/H] gradients over time. The values plotted here are the slopes to the best linear fit, as the exponential fit does not apply to every halo (e.g., Au-10G and Au-17G). The horizontal axis for the first two rows (last row) shows the \fehgradrinf~(\fehgradrnow)~on the left and the \fehgradeinf~(\fehgradenow)~on the right. The vertical axes for the top panels indicate the present-day metallicity gradients, i.e., \fehgradrnow(left) and \fehgradenow(right), while the middle and bottom panels quantify the percentage change between the pre-merger and the present-day [Fe/H] gradients. Here, a value closer to zero indicates little difference between the pre-merger versus post-merger (radial or energy) metallicity gradient. The blue-shaded regions on the right panels show the range of percentage change in the radial [Fe/H] gradient. 
To identify the strength of the trends, we also calculated their Pearson Correlation Coefficients (CC) where a value of 0.5 to 1 means strong correlation, 0 means no correlation, and -0.5 to -1 means strong anti-correlation.  

The top-left panel highlights a moderately negative correlation between the slope of the radial metallicity gradient today and at infall, with a CC of -0.46. That is, progenitors that had steeper \fehgradrinf~tend to have shallower \fehgradrnow. At face value, this seems counter-intuitive; one might assume that systems that had steeper gradients at infall had more of the gradient to "preserve", and would similarly have steeper gradients at present-day. But we argue that this is not too surprising given the range in slopes for the pre-merger vs post-merger trends, as well as the physical properties of these progenitors which we discuss later in Section \ref{sec:grad_v_props}. On the other hand, the trends with energy (top-right panel) generally show a different behavior. The CC for \fehgradenow~vs \fehgradeinf~ is -0.02, although this seems to be largely driven by Au-10G. Without this halo, there is a much clearer positive correlation, with a CC of 0.86. 
This trend in \fehgradenow~vs \fehgradeinf~highlights the important nature of energy in redistributing stars, such that systems with steeper \fehgradeinf~also have steeper \fehgradenow.

The middle-left panel shows that the percentage change in the radial metallicity gradient is highly anti-correlated with \fehgradrinf~with a CC of -0.79. That is, there is a greater change in the metallicity gradient for systems that originally had steeper \fehgradrinf. In fact, this anti-correlation is even more apparent if we consider instead the absolute difference between the present-day and infall slopes, with a CC of -0.99.  We note, however, that the range in \fehgradrinf~is -0.100 to -0.020 dex/kpc while the range in \fehgradrnow~is -0.008 to -0.001 dex/kpc, the former having much more significant gradients than the latter. It would therefore follow that the largest change would occur for the steepest gradients as those with shallower gradients at infall have less of a gradient to be washed out. 
This is similarly the reason for the strong anti-correlation in the percentage change in energy metallicity gradient with \fehgradeinf~(middle-right panel), where the CC is -0.94. In addition, the percentage change for the energy metallicity gradient spans a wider range and reaches lower values compared to that of the radial trend, highlighted as the blue region. Some GES progenitors change energy metallicity gradients very minimally (e.g., Au-27G), while some get almost completely washed out (e.g., Au-10G). This wide range together with the strong anti-correlation with \fehgradeinf~emphasizes further that looking at the metallicity gradient with respect to energy is a more viable space to explore the past of the GES. 

Lastly, from the bottom-left panel, we see that systems with shallower \fehgradrnow~tend to have the largest percentage change in their gradients, with a CC of 0.87. The trend in the percentage change against \fehgradenow~is skewed by Au-10G (CC = 0.25) but without this halo, there is a stronger negative correlation with a CC of -0.70 where progenitors with the shallowest \fehgradenow~underwent the least change. These correlations between present-day gradients and their percentage changes are a potential window to tracing back the original metallicity gradient that the GES progenitor fell in with. 

In the next section, we investigate how the properties of the GES merger in the simulations drive the differences and trends that we see with metallicity.

\subsection{[Fe/H] Gradient vs GES properties}
\label{sec:grad_v_props}

\subsubsection{GES radial [Fe/H] gradient}
\label{sec:grads_feh}

\begin{figure*}
\includegraphics[width=0.9\textwidth]{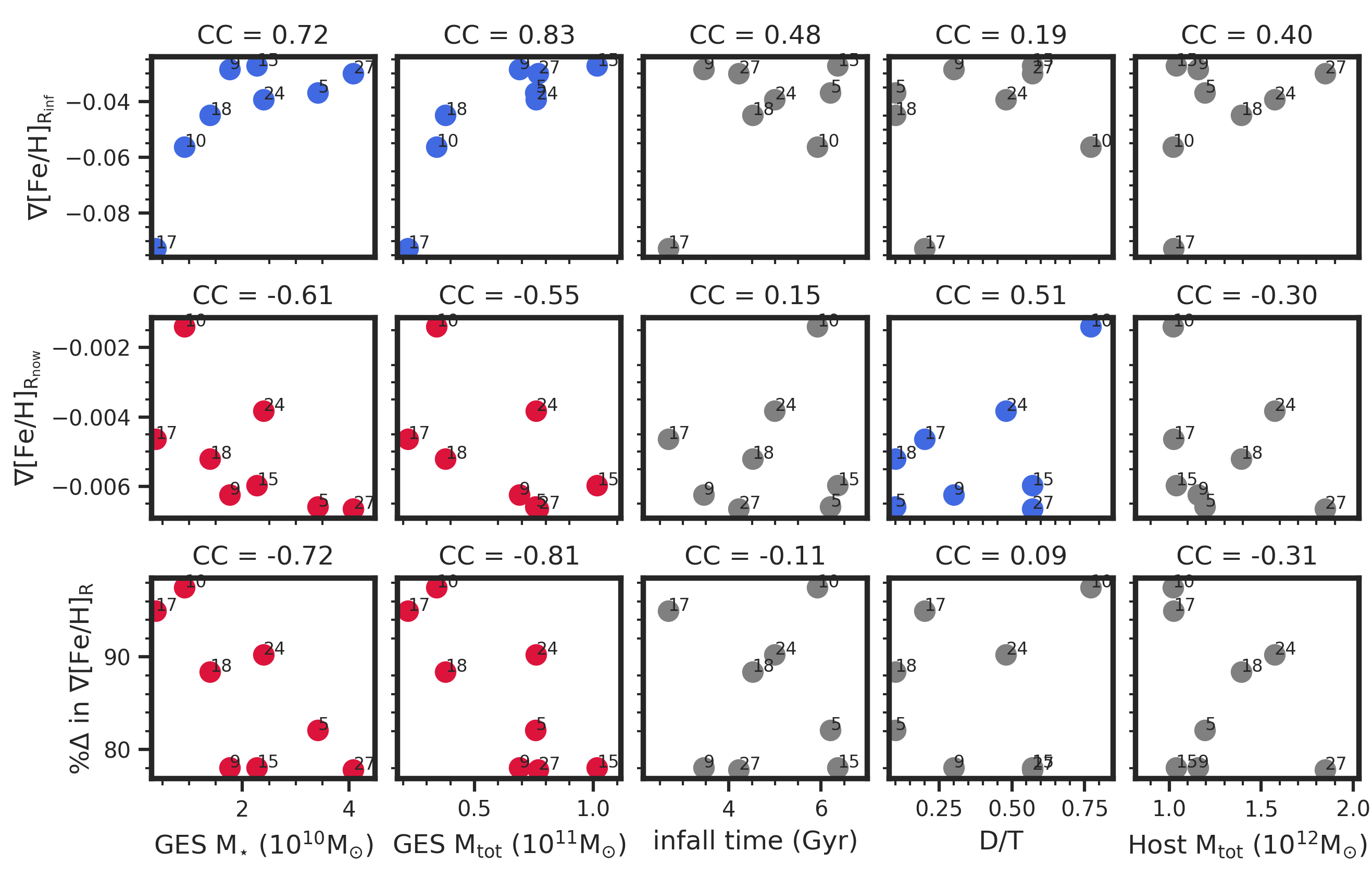}
 \caption{\textbf{[Fe/H] gradients with radial distance as a function of merger property.} Radial metallicity gradient at infall (\fehgradrinf), at present-day (\fehgradrnow), and the percentage difference (with respect to infall) between the current and infall radial metallicity gradients from top to bottom, against the GES infall stellar mass ($\rm 10^{9} M_{\odot}$), GES infall total mass ($\rm 10^{11} M_{\odot}$), infall time (Gyr), GES progenitor disk-to-total mass ratio, and the host peak total mass (in $\rm 10^{12} M_{\odot}$), from left to right. The Pearson correlation coefficients (CC) are indicated at the top of each sub-panel, and the trends that are strongly correlated (CC $>$ 0.5) and strongly anti-correlated (CC $<$ -0.5) are colored blue and red, respectively.  }
 \label{fig:grads_v_props_feh}
\end{figure*}

Figure \ref{fig:grads_v_props_feh} explores the trends in \fehgradrinf, \fehgradrnow, and the percentage change in the radial metallicity gradient, $\rm \nabla [Fe/H]_{R}$, against the GES infall stellar mass, GES infall total mass, GES infall time, GES $D/T$, and host total mass at present day from left to right. We specifically compare against these merger properties as we presume that they would have a significant effect on the metallicity gradient. We similarly calculate the CC for these trends which are indicated on the top of each subpanel. The markers are colored blue for positively correlated properties ($\rm CC \geq 0.5$), gray for zero to low correlation ($\rm -0.5 < CC < 0.5$), and red for negatively correlated properties ($\rm CC \leq -0.5$).  Though we report these CC's, we do caution that our small sample prohibits us from making strong claims based on these numbers. We use these mostly as a guide for our physical interpretation.

All of the radial gradients (and percentage change in gradients)
have strong correlations with the GES stellar mass and total mass. The most massive GES progenitors have shallower gradients at infall. These, in turn, have the steepest gradients at present-day and the smallest change in the radial metallicity gradient. 

The relationship between $\rm \nabla [Fe/H]_{R}$ and the mass of a galaxy is still highly debated, with different observational works finding these two properties to be correlated \citep[e.g.,][]{belfiore17,franchetto21} and anti-correlated \citep[e.g.,][]{lutz21,ho15} in the same mass regime as the GES progenitor. We discuss this in greater detail in Section \ref{sec:metgradobs}. We note that these works have also used various tracers i.e., stars, gas-phase metallicity, and atomic gas (H\textsc{i}). We can also make a direct comparison with other simulations. For example, \citet{khoperskov23} used the HESTIA cosmological simulations as well as 1000 N-body simulations to answer the same question we are asking---\textit{what was the GES progenitor's metallicity gradient?} Their Figure 6 shows that less massive progenitors have a larger scatter in \fehgradrinf, reaching steeper gradients at infall, similar to what we are finding in this work. We posit that this strong correlation between the \fehgradrinf~and the GES stellar and total mass is therefore a viable scenario. 

With the relationship between the \fehgradrinf~and GES stellar and total mass established, it is easier to make sense of the trends in the \fehgradrnow~and the change in $\rm \nabla [Fe/H]_{R}$. The less massive progenitors have shallower potential wells and so their stars can be more easily stripped. The effect of washing away an existing metallicity gradient would then be more pronounced for these systems, and they would thus end up with the shallowest \fehgradrnow~as we see for the second row in Figure \ref{fig:grads_v_props_feh}. In addition, steeper gradients are more sensitive to changes along the distance direction compared to shallower gradients, merely because shallow gradients are going to stay shallow even after reshuffling the stars in distance. 
Therefore, the fact that the least massive GES-like progenitors that had steeper \fehgradrinf~exhibit the largest change in radial metallicity gradient is a logical course of evolution.  Another potential mechanism at play is that the more massive progenitors (e.g., Au-5G and Au-27G) are subject to more dynamical friction and therefore change to their orbit. They plunge deeper into the galaxy in a shorter timescale, and as they lose stars in this process, this creates a steeper gradient at present-day compared to the less massive progenitors with farther apocentres.

The infall times and host total mass do not correlate with the radial metallicity gradient measurements, while the $D/T$ is on the edge of a positive correlation with \fehgradrnow. This may hint at how mergers with a larger contribution from a disk tend to have shallower gradients today. However, because this is very close to our cut (i.e., the CC is different by only 0.01) and Au-10G highly skews this relationship, we caution that this is a tenuous trend.

\subsubsection{GES energy [Fe/H] gradient}
\label{sec:grads_energy}

\begin{figure*}
\includegraphics[width=0.9\textwidth]{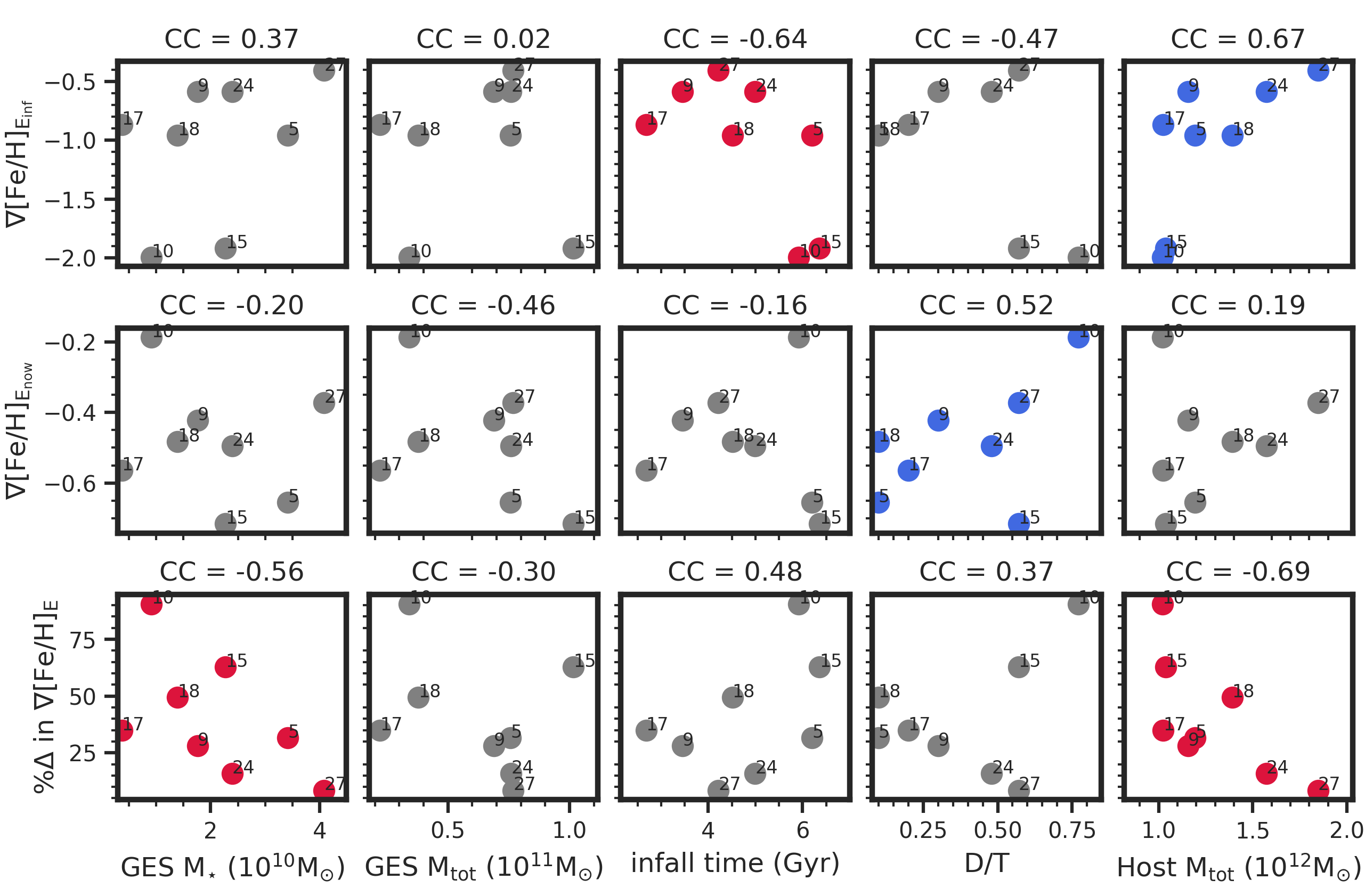}
 \caption{\textbf{[Fe/H] gradients with orbital energy as a function of merger property.}  Metallicity gradient with respect to orbital energy at infall (\fehgradeinf), at present-day (\fehgradenow), and the percentage difference (with respect to infall) between the current and infall energy metallicity gradients from top to bottom. The columns are arranged similar to those in Figure \ref{fig:grads_v_props_feh} with the points colored blue to indicate strong correlation (CC $>$ 0.5) and red for strong anti-correlation (CC $<$ -0.5)}
 \label{fig:grads_v_props_BE}
\end{figure*}

In addition to the radial metallicity gradient, we also investigate the [Fe/H] gradient with respect to energy at infall (\fehgradeinf), at present-day (\fehgradenow), and the percentage difference between them (\% $\Delta$ in $\rm \nabla [Fe/H]_{E}$), against different merger properties in Figure \ref{fig:grads_v_props_BE}. We compare to the same properties as in Figure \ref{fig:grads_v_props_feh} and similarly color the markers based on their Pearson CC i.e., blue for positively-correlated, red for negatively-correlated, and gray for no correlation. 

Interestingly, the same exact panels are not necessarily illuminated between Figures \ref{fig:grads_v_props_feh} and \ref{fig:grads_v_props_BE}. \fehgradeinf~and \fehgradenow~do not have strong correlations with the GES mass, unlike the case for the radial gradients. Only the percentage difference in $\rm \nabla [Fe/H]_{E}$ exhibits a trend with GES (stellar) mass with a CC of -0.56. Similar to the case for the radial gradient, the more massive systems exhibit the least change in the energy gradient between infall and present day because their stars are more bound, driving this anti-correlation.
The fact that this trend only holds for the stellar mass and not the total mass may seem curious, especially because the energy of a star would be affected most by the total mass of the galaxy. 
When we do check against the \textit{peak} stellar and total mass, both have strong anti-correlations with the percentage change, having CC of -0.54 and -0.59, respectively. We also note that Au-15G falls in as a double-system and seems to be an outlier; removing this GES progenitor further strengthens the anti-correlation for the percentage difference in $\rm \nabla [Fe/H]_{E}$ vs GES stellar mass and total mass, with CC of -0.66 and -0.70, respectively. We again emphasize that the percentage change encompasses a huge range for the energy metallicity gradient which goes from $\sim$9-91\% compared to that of the radial metallicity gradient which are much higher at $\sim$78-98\%. 

The infall time is negatively-correlated (CC = -0.64) with \fehgradeinf~and slightly positively-correlated (CC = 0.48) with the change in the energy metallicity gradient. One possible explanation for the anti-correlation between infall time and \fehgradeinf~is that GES-analogs accreted more recently have had more time to accrete lower mass systems as they have spent more time as a ``central'' system \citep{taibi22}. 
These smaller accreted systems would have relatively lower metallicities compared to the main GES galaxy \citep{kirby13} and due to their lower mass, they would be destroyed more easily and dispersed into the GES stellar halo. 
This can altogether cause a steepening in the energy metallicity gradient. Some other effects are most likely at play here but we note that the hierarchical formation of galaxies would have a non-negligible effect on this anti-correlation that we see. The mildly positive correlation between the percentage difference and the infall time is a natural consequence of steeper gradients being washed out the most.

The $D/T$ is yet again on the cusp of a positive correlation with the present-day energy metallicity gradient at CC = 0.52, similar to the case for the radial trend. The fact that the same correlation exists for both radial and energy gradients against $D/T$ lends more weight to this property in terms of the fate of GES which we discuss in more detail in Section \ref{sec:discussion}. A stronger presence of a disk component in the GES progenitor roughly results to a shallower metallicity gradient at present-day. Interestingly, Au-10G and Au-15G visually appear to be outliers here, and removing them results in strong relationships against $D/T$, i.e., CC = 0.95 for \fehgradeinf, CC = 0.70 for \fehgradenow, and CC = -0.92 for the percentage change. Though the $D/T$ and metallicity gradient for Au-15G may be unreliably measured as it is falling in with another system, it is less clear why Au-10G would be an outlier. It is interesting to note however that Au-10G is the diskiest of the progenitors in our sample, followed by Au-15G and Au-27G.


The host's present-day total mass is positively correlated with \fehgradeinf~with CC = 0.67, 
in that the more massive Milky Way-like hosts in Auriga tend to have accreted GES progenitors that exhibit shallower \fehgradeinf. However, this is unlikely a consequence of the GES total mass (i.e. more massive hosts typically accrete more massive GES systems), which shows little correlation with \fehgradeinf~(top row, second column of Figure \ref{fig:grads_v_props_BE}). It is puzzling that a GES infall property would be correlated with a present-day host property. This may suggest a potential link between the \fehgradeinf~and the environment the GES progenitor is in (perhaps a result of assembly bias), though the exact reason would have to be explored further. We do caution that this is largely due to Au-10G and Au-15G and it is possible that the correlation is mere coincidence, and not astrophysically driven.  
With this strong positive correlation for \fehgradeinf~vs host mass, it follows that the percentage difference in $\rm \nabla [Fe/H]_{E}$ exhibits a strong negative correlation as the steepest gradients would experience the greatest change. 
\\
\\
In this Section, we explored the presence and destruction of the GES metallicity gradient against different merger properties.
The GES infall stellar and total masses are  correlated with the initial radial [Fe/H] gradient, while the infall time and present-day host total mass are correlated with the initial energy [Fe/H] gradient. The $D/T$ for the GES progenitor is consistently correlated with the present day [Fe/H] gradient in both radial and energy space, where diskier GES tend to end with shallower gradients today. Lastly, we have multiple avenues to potentially infer the percentage change in the [Fe/H] gradient with respect to the GES galactocentric radius 
and to the progenitor orbital energy given the GES mass, its infall time, and the Milky Way total mass today. The percentage change has a larger, more dynamic range (9-91\%) for energy than radius (78-98\%), highlighting that energy is a more conserved quantity, though we note that the radial trends are easier to measure. 


\section{Discussion}
\label{sec:discussion}

We have shown both qualitatively and quantitatively how the metallicity gradient in GES-like progenitors in Auriga evolve through time---from infall to the present-day. These GES progenitors have a negative metallicity gradient that existed before falling into their respective hosts, and this negative gradient persists both against radius and energy to today, though as a much weaker signature. More importantly, we found that different physical properties such as GES stellar and total mass, infall time, and host total mass could probe the changes in these gradients, which could go from barely having changed (at 9\% difference) to washing away almost completely (at 98\% difference). This result suggests that accurately knowing the GES stellar and total mass, its infall time, and the Milky Way total mass is a potentially powerful probe into the chemical cartography of GES's past. But before we are able to jump back in time, we first discuss how the metallicity gradients in the simulations compare to observations of similar-mass galaxies, and how to approach reconstructing the GES cartography from real Milky Way data.

\subsection{Metallicity gradient in observations}
\label{sec:metgradobs}

To ensure a wider scope for comparison, we look at both gas-phase and stellar metallicity gradients in the observations. Comparing to gas-phase metallicity is especially important at higher redshifts where emission from gas is easier to measure than the continuum from stars. We recognize that this is not a perfect comparison. Stars are affected by radial migration, so the metallicity gradient of older stellar populations would be more blurred out already compared to that of younger stars forming out from the gas \citep{frankel18}. Gas dynamics may affect the gas-phase metallicity gradient and therefore not necessarily reflect the distribution of stars, and the sample itself would be limited to those with measurable emission (and are therefore largely star-forming). In addition, studies of metallicity gradients focusing on systems with stellar masses similar to GES (i.e., $\sim10^{9}~\rm M_{\star}$) are not largely sampled or are typically on the cusp of statistically large surveys such as MaNGA \citep{bundy15} and CALIFA \citep{sanchez12}. Most importantly, we are only able to compare to the \textit{radial} metallicity gradient, which is more easily measured than the \textit{energy} metallicity gradient in observations. Nonetheless, to anchor our results in the simulations to the real universe, it is necessary to compare to observations of similar mass systems both at high and low redshifts.


\subsubsection{High redshift galaxies}
We now look at high redshift galaxies at $1<z<3$ to more accurately sample a GES-like progenitor before it merged with the Milky Way. Gravitational lensing has been especially useful in probing the metallicity gradients of high redshift galaxies, providing us a spatially-resolved view of galaxies that are otherwise too small to observe. In addition, the samples from gravitationally-lensed works are unbiased, as the alignment of a lensing galaxy in the foreground and the lensed galaxy in the background is random.

To this effect, studies from the Grism Lens-Amplified Survey from Space (GLASS) survey using the Hubble Space Telescope have been helpful in contextualizing our results in contrast to high redshift observations. \citet{wang17} looked at the gas-phase metallicity of star forming galaxies in the $1<z<2$ regime that were serendipitously magnified by the cluster MACS1149.6+2223. 
They report a diversity in morphologies and gradients (i.e., negative, flat, and positive) of galaxies that have a mass range of $10^{8-9.5}~M_{\odot}$. 
Interestingly, they similarly find shallower radial metallicity gradients at the higher mass end of this range (as we see in our Figure \ref{fig:grads_v_props_feh}). They posit that such a correlation may result from massive galaxies having more evolved disks with star formation happening at all radii, flattening the metallicity gradient. This is similar to what \citet{curti20} found using the KMOS Lensed Emission Lines and VElocity Review (KLEVER) Survey. However, this does not necessarily mean that the diskiest galaxies have the shallowest metallicity gradients (as we see for Au-10G which seems to show the opposite). Both the mass and the $D/T$ should be high for a GES-like galaxy in Auriga to have a shallower radial and energy metallicity gradient at infall, such as the case for Au-24G and Au-27G. It is important to note though that the GES progenitor likely has a lower (stellar) mass than Au-24G and Au-27G based on observations \citep[e.g.][]{callingham22,lane22,carrillo24}. Other works from the GLASS survey have also shown that galaxy interactions can affect the resulting metallicity gradient. \citet{jones15} found a shallower slope for a lensed galaxy at $z=1.85$ compared to isolated galaxies at the same redshift because of interactions with a neighboring system. This is similar to what we see in Figure \ref{fig:premerge} for Au-15G, which falls into the host's virial radius with another subhalo.

The James Webb Space Telescope (JWST) has further revolutionized our view of the spatially-resolved high redshift universe. Initial works from the MSA-3D survey \citep{ju24} using JWST NIRSpec estimated the radial metallicity gradients for galaxies with stellar masses $8.69 \leq log_{10}(M_{\star}/M_{\odot}) \leq 11.20$ at $0.5<z<1.7$ without the aid of gravitational lensing. At the redshift that we are investigating (i.e., $z>1$), their sample of galaxies have negative and flat radial metallicity gradients, with similar values to what we find in the simulations. They also find that galaxies with higher stellar masses tend to have more negative gradients. Although this is contrary to the positive correlation that we see in Figure \ref{fig:grads_v_props_feh}, we do note that this trend seems to be driven by systems that have \mstar$>10^{10}M_{\odot}$ in their sample. Overall, the GES progenitors in Auriga exhibit reasonable and consistent radial metallicity gradients with observations of galaxies at similar mass and redshift ranges. 



\subsubsection{Low redshift galaxies}

As the GES is a destroyed dwarf satellite of the Milky Way, it is important to compare the metallicity gradients of surviving satellites in the Milky Way and the Local Group at large. 
Fortunately, stellar metallicity gradient measurements are available for these systems, making them more directly comparable to  study. \citet{kirby11} measured the stellar radial metallicity gradient for 8 Local Group dwarf galaxies, which they found to have largely negative gradients similar to what we found in the simulations. \citet{leaman13} further showed that the dwarf irregulars (dIrr) such as the LMC, SMC, and WLM (an isolated dIrr) have shallower gradients compared to dwarf spheroidals (dSph) like Fornax and Sculptor. They propose that this is because the dIrr galaxies have higher angular momentum compared to the dSph galaxies, which prevents chemically-enriched gas from getting funneled easily into the central parts, therefore flattening the metallicity gradient. From the comparison of \fehgradeinf~to the $D/T$ in Figure \ref{fig:grads_v_props_BE} (and assuming that Au-15G and Au-10G are outliers), we can generally see the same correlation observed in the Local Group i.e. our diskier systems tend to have shallower gradients. However, we do note that this relationship is more pronounced in \fehgradeinf~than in \fehgradrinf~and the latter is what we could more easily compare to in the observations. \citet{taibi22}
extended their analysis to 30 Local Group dwarfs in order to more statistically understand what drives the stellar metallicity gradient. 
They find that the radial metallicity gradient is shallower for more luminous and therefore more massive systems, similar to what we find in Auriga. We note that their sample of galaxies has a lower mass range than what we are looking at (i.e. $M_{\star} \sim 10^5$ to $10^9$ $\rm M_{\odot}$) although there is overlap at the higher mass end. Among the Local Group dwarfs that match the radial metallicity gradients in the simulations, two are dIrr i.e., SMC and IC 1613, and one is a dSph, Antlia II, which is also interestingly on the lower mass end of the sample. They find, however, that when the radial metallicity gradient is measured and scaled to the half-light radius, this correlation with mass disappears. These studies show that what causes the slope of the stellar metallicity gradient is not so straightforward, and, as we have seen in the simulations, could be affected both by the mass and the morphology of the system. Nonetheless, the general trends match in that the most massive satellites in the Local Group---the ones more comparable to GES---have the shallowest gradients and have a similar range to the GES analogues in the simulations.

It is worth noting that \citet{grand19metals} has explored how metals are redistributed in the Milky Way-like central galaxies in Auriga and found that fountain flows are able to change low angular momentum metal-rich material into high angular momentum material, therefore flattening the metallicity gradient. Although investigating the exact cause of the metallicity gradient at GES-mass scales
is beyond the scope of this paper, it is worth pointing out the importance that angular momentum likely plays in shaping the metal distribution in galaxies, destroyed or intact, central or satellite.

For completeness, we also compare to nearby (non-Local Group satellite) galaxies, with the caveat that such samples of galaxies would be the least comparable to the GES-like systems in this work. This is because these samples tend to have central galaxies with higher masses than our sample.  \citet{ho15} tried to alleviate this in their work by supplementing CALIFA IFU data with observations of local galaxies with the Wide-Field Spectrograph at the ANU 2.3m Telescope. This covers the mass range of $M_{\star} = 10^{8-9} M_{\odot}$, which we can compare to our GES sample in Auriga. They found that the gas-phase metallicity gradient flattens with increasing mass at this regime (see their Figure 9) although when scaled to the half-light radius, any trends with stellar mass disappears, similar to what \citet{taibi22} found for the stellar metallicity gradient in the Local Group. \citet{belfiore17} used the SDSS-IV MaNGA IFU survey to explore the trends in the gas-phase metallicity gradients of 550 nearby galaxies. They find that the gradient actually steepens with stellar mass, the opposite of what we find. However, in their lower-mass regime, which is comparable to our sample (at $M_{\star}\sim 10^{9}$), they measure shallow negative or flat gradients. Although the comparison is harder to make with these local galaxies because of the aforementioned constraints on the mass range, the metallicity gradients measured for GES in the simulations are largely in good agreement with the observations. 


\subsection{Applications to the real Milky Way}

In this work, we have investigated the pre- and post-merger metallicity gradients of GES analogues in Auriga to ultimately rewind time and reconstruct the GES progenitor from the real data. Although we have determined that certain properties such as stellar and total mass, infall time, and host mass are a window to the initial metallicity gradient of GES, our analysis was done fully knowing which stars came from GES and which did not. This is simply not the case for the observations, especially for the halo and even more so for the GES which spans a huge range in different dynamical and chemical spaces, therefore being prone to contamination. Of course, it is best to have both a pure \textit{and} complete sample that accurately represents the real metallicity distribution of GES over a large range of Galactocentric radii. However, purity comes at the expense of completeness (and vice versa) and multiple works have shown the interplay of these two quantities with different GES selection methods \citep[e.g.][]{buder22, lane22,carrillo24}. For example, \citet{carrillo24} benchmarked observational selection methods of GES using the Auriga simulations with GES-like systems, and found that a selection of GES stars using the action diamond method gives the most accurate stellar mass estimate for the progenitor, even if it does not give the purest or most complete sample. It is therefore necessary to test from the simulations which levels of purity and completeness are \textit{good enough }to still accurately reconstruct the GES progenitor metallicity gradient in the observations. In addition, our analysis could be extended with the use of mock data that takes into consideration the selection function of Milky Way surveys. For example, mock \textit{Gaia} data from cosmological hydrodynamical simulations have been produced such as \textit{ananke} \citep{sanderson20} using FIRE-2 simulations \citep{wetzel16} and \textit{AuriGaia} using Auriga simulations \citep{grand18}. Other works have extended this effort and modeled the combined selection function of \textit{Gaia} and other spectroscopic surveys such as APOGEE (e.g., \citealt{nikakhtar21}) and DESI (e.g., \citealt{auridesi24}) in hopes of better translating between simulation and observational results.   

Another consideration to make when applying this analysis to the observed Milky Way data is the need for a large sample of stars in the halo that reach larger Galactocentric distances. The present-day metallicity gradient of GES is most likely substantially shallower than its pre-merger metallicity gradient, as we have demonstrated in Figure \ref{fig:postvpre_grad}. To probe stars that were mostly in the outskirts of the progenitor before it fell into the Milky Way, we would need observations at $>$50 kpc (see Figure \ref{fig:postmerge}) to more accurately reconstruct the progenitor metallicity gradient using a representative sample. Thankfully, many next-generation large spectroscopic surveys such as DESI Milky Way Survey \citep{cooper23} and WEAVE \citep{jin24} are ideal for tackling this problem, providing velocities and metallicities for distant stars that will enable us to get metallicity gradients against distance and energy. In fact, using RR Lyrae stars that extend up to 100 kpc in DESI, \citet{medina25} measured a metallicity gradient for a selection of GES stars to be -0.005 dex/kpc. This seems to be consistent with our measurements in the simulations for GES-like progenitors (see Figure \ref{fig:grads_v_props_feh}), where we can infer mass of $\sim 2 \times 10^{9} \rm M_{\odot}$ in stellar mass and $\sim 6 \times 10^{10} \rm M_{\odot}$ in total mass for the GES in observations.  

Lastly, a key result from this work is that by accurately knowing the GES stellar and total mass, its infall time, and the total mass of the Milky Way---properties that we can measure from observations---we can estimate \textit{how much} the radial and energy metallicity gradients have changed over time as we have shown in Figures \ref{fig:grads_v_props_feh} and \ref{fig:grads_v_props_BE}. It is therefore imperative to accurately determine these properties as well to have better combined constraints in reconstructing the distribution of metallicity across the GES progenitor. To date, stellar mass estimates for GES have a huge range between $1.5 \times 10^{8}~M_{\odot}$ \citep{lane23} to $5 \times 10^{9}~M_{\odot}$ \citep{vincenzo19}. The total mass estimates are even harder to determine, but have a range from $10^{10 -11}~M_{\odot}$ \citep{belokurov18,das20}. Although these stellar and total mass ranges are well within the GES progenitors in Auriga, narrowing them would be necessary for better constraints on the progenitor. 
The GES progenitors in Auriga have a wider range of infall times compared to the GES in the observations \citep[e.g.][]{helmi18,bonaca20} but these estimates do fall within our range in the simulations, making this a viable parameter space in constraining the infall metallicity gradient. Lastly, the present-day total mass of the host seems to show a strong negative correlation with the energy metallicity gradient, although the exact link is yet to be explored and beyond the scope of this paper. Nonetheless, better constraints on the total mass of the Milky Way are needed for purposes that extend far beyond the goal of this work, and this requires careful modeling from the Galaxy rotation curve as pointed out by \citet{oman24}.

\subsection{Caveats}

Although we explored in great detail the metallicity gradient of possible GES progenitors, we acknowledge that this is but one suite. We use the Auriga simulations because they have a largely-tested sample of GES-like progenitors  \citep[e.g.,][]{fattahi19,callingham22,orkney23,carrillo24}, but it is worth exploring other cosmological hydrodynamical simulations as well 
to confirm the robustness of our results. For this, it would be better to compare the percentage difference in the pre- vs post-merger metallicity gradients between different simulations. The percentage difference would presumably be less sensitive to the galaxy formation and chemical enrichment recipes, and therefore better probe the dynamical evolution of GES stars in the same way we are doing in this study.

In addition, Auriga has 30 Milky Way-like galaxies at Level 4 (with a baryonic mass resolution of $5 \times 10^{4} \rm M_{\odot}$), and from this sample, we selected eight galaxies that had a GES-like progenitor from the work of \citet{fattahi19}. We acknowledge that this is a small sample and some trends are skewed by certain outlier progenitors such as Au-10G and Au-15G. Therefore, a larger sample is necessary to more statistically determine the validity and strength of the correlations between metallicity gradient and GES properties. 

With regards to the radial and energy metallicity gradients, we find that the radial metallicity gradient undergoes a much larger change compared to the energy metallicity gradient. The former is also generally shallower and therefore less discerning than the latter. We therefore find similar trends as \citet{khoperskov23} which used the energy metallicity gradient to determine the GES progenitor's metallicity gradient from APOGEE data. However, we do note that the energy is dependent on the Milky Way potential model and input observables (i.e., distance, velocities, positions) while the radial metallicity gradient is a more direct measurement. In addition, although we perform a linear fit, it does not fully capture the metallicity gradient at infall for some of our progenitors. In fact, some of them are better fit by an exponential function as shown in Figure \ref{fig:premerge}, where the gradient climbs up more drastically towards the center. However, not every system is describable with an exponential fit at infall and all halos are best fit with a line at present-day. Comparing a linear fit at infall vs a linear fit at present-day thus makes the interpretation of the percentage change more tractable. For the same reason, we have excluded the inner most region in fitting to the energy metallicity gradient because of a slight plateau (or even a turnover) as seen in Figures \ref{fig:premerge} and \ref{fig:postmerge}.

Lastly, exploring the gas content of the GES progenitors in Auriga in detail is outside the scope of this work, but it is worth investigating the interplay of the gas fraction, diskiness, and the resulting metallicity gradient for the GES progenitor. 




\section{Conclusion}
\label{sec:conclusion}
We carried out this study to understand the pre-merger and post-merger picture of the GES system using the Auriga hydrodynamical cosmological zoom simulations. These GES-like progenitors in Auriga have a variety of morphologies and diskiness (Figure \ref{fig:ges_xz}, see also \citealt{orkney23}) which is informative in finding analogs at $z\sim2$. We specifically investigated the stellar metallicity gradient---its presence and destruction. In doing so, we arrive at the following results:  

\begin{itemize}  
\item Negative metallicity gradients exist at infall against both the GES-centric radius and the orbital energy of stars in the GES progenitor (see Figure \ref{fig:premerge}), that range between -0.0925 to -0.0284 dex/kpc and -1.9915 to -0.4081 dex/$\rm 10^{-5} km^{2}s^{-2}$, respectively. The gradients we find in the simulations are in line with observations of similar-mass galaxies (Section \ref{sec:metgradobs}). We also find that interactions with a nearby subhalo can lead to stretching out the gradient, as seen in Au-15G. This ultimately happens to all the progenitors once they get accreted onto their host galaxies. 
\item Nevertheless, there remains a negative metallicity gradient at present-day for the radial and energy trends (see Figure \ref{fig:postmerge}) and both are shallower (by an order of magnitude) than at infall with a range of -0.0066 to -0.0014 dex/kpc and -0.7149 to -0.1872 dex/$\rm 10^{-5} km^{2}s^{-2}$, respectively.
\item At the present-day, \textit{where} the GES stars get deposited in the host galaxy is a valuable probe to its origin (see Figure \ref{fig:postmerge}). At the same galactocentric radius, lower [Fe/H] stars generally come from larger GES-centric radii at infall. In addition, at the same metallicity, stars deposited at larger radii also typically come from larger radii in the progenitor. We find a similar trend for the energy metallicity gradient. 
\item Progenitors with the steepest infall \textit{radial} metallicity gradients tend to be the shallowest at present-day, while those with the steepest infall \textit{energy} metallicity gradients remain the steepest today (see Figure \ref{fig:postvpre_grad}).  From the same figure, we see that the radial metallicity gradient undergoes a much larger percentage change from 78-98 \%, while that for energy has a more dynamic range from 9-91 \%. The progenitors that conserve their energy metallicity gradients the most tend to be the most massive. 
\item Lastly, we quantify the correlation between infall, present-day, and percentage change in the radial (Figure \ref{fig:grads_v_props_feh}) and energy (Figure \ref{fig:grads_v_props_BE}) metallicity gradients against GES progenitor and merger properties. Interestingly, different properties are highly correlated for the radial trend versus the energy trend. Exploring both spaces would therefore be better for inferring the GES metallicity gradient at infall, especially as we get better constraints for the total and stellar mass of GES, its infall time, and the total mass of the Milky Way. 
\end{itemize}

Altogether, our exploration provides a glimpse into the past of this large building block of our Galaxy through leveraging the data that we \textit{do} have at present. Amidst a list of caveats and careful considerations for applying this to real Milky Way data, we emphasize that examining the connection between the chemistry, positions, and energies of accreted stars with the progenitor and host properties is a necessary step to reconstruct this most significant merger before its eventual demise.

\section{Acknowledgements}
AC and AD acknowledge support from the Science and
Technology Facilities Council (STFC) [grant numbers 
ST/T000244/1 and ST/X001075/1] and the Leverhulme Trust. AD is supported by a Royal Society
University Research Fellowship. RG acknowledges financial support from an STFC Ernest Rutherford Fellowship (ST/W003643/1). AF acknowledges support by a UKRI Future Leaders Fellowship (grant no. MR/T042362/1) and a Sweden's Wallenberg Academy Fellowship. FF is supported by a UKRI Future Leaders Fellowship (grant no. MR/X033740/1).   

This work used the DiRAC@Durham facility managed by the Institute for Computational Cosmology on behalf of the STFC DiRAC HPC Facility (www.dirac.ac.uk). The equipment was funded by BEIS capital funding via STFC capital grants ST/K00042X/1, ST/P002293/1, ST/R002371/1 and ST/S002502/1, Durham University and STFC operations grant ST/R000832/1. DiRAC is part of the National e-Infrastructure.

\section{Data Availability}
The Auriga Level 4 simulations are available at \url{https://wwwmpa.mpa-garching.mpg.de/auriga/}. 



\bibliographystyle{mnras}
\bibliography{aaa_ges} 




\appendix

\section{Post-merger gradients of other GES-like halos in Auriga}
\label{app:other_halos}

\begin{figure*}
\includegraphics[width=\textwidth]{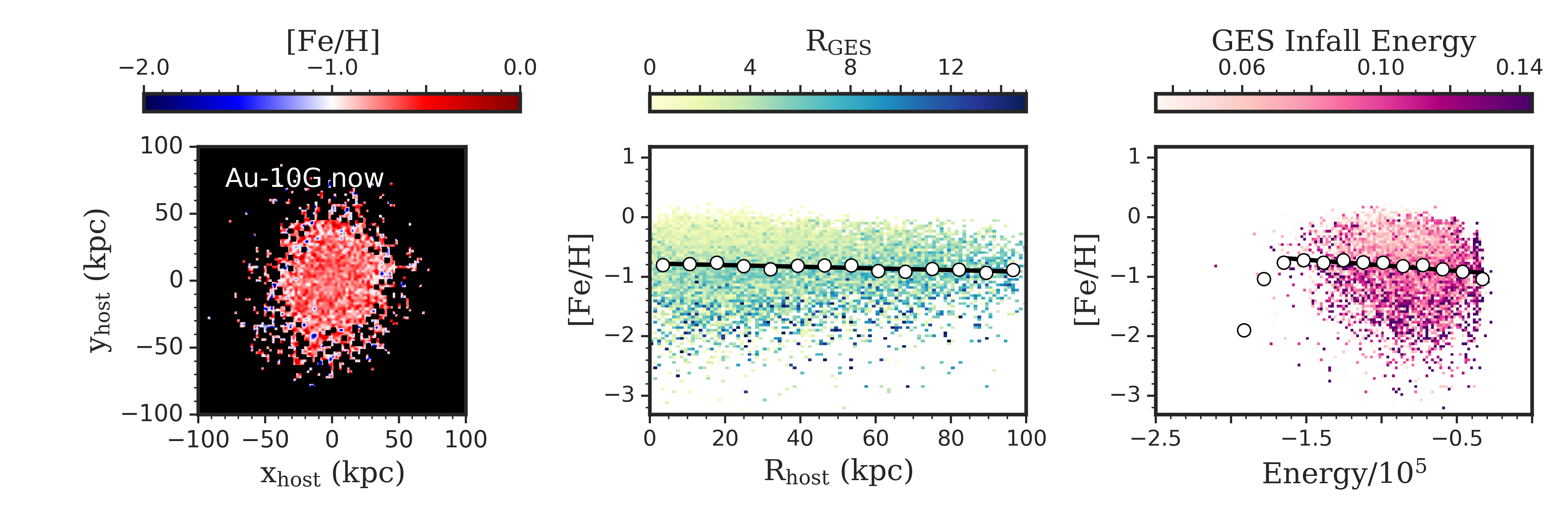}
\includegraphics[width=\textwidth]{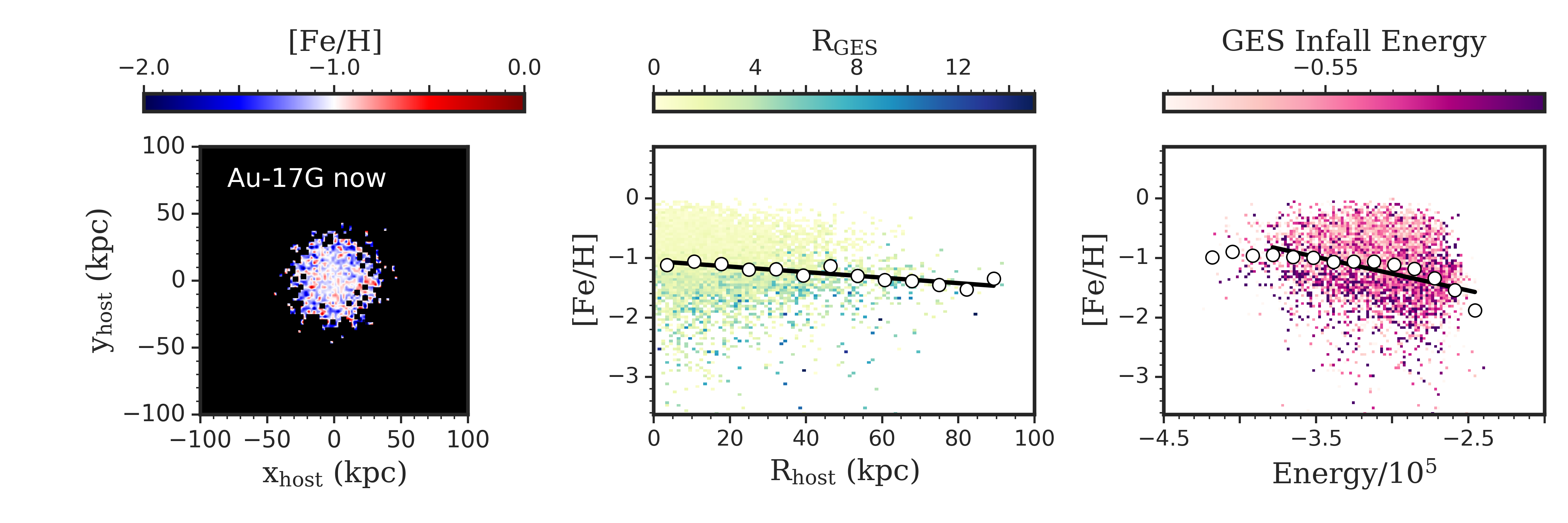}
 \caption{Figure is similarly arranged as Figure \ref{fig:postmerge} but highlight two other halos, Au-10G and Au-17G, whose mapping between infall time versus present-day properties are less straightforward compared to that of Au-5G and Au-15G. } 
 \label{fig:postmerge_app}
\end{figure*}

Figure \ref{fig:postmerge_app} shows two other halos, Au-10G and Au-17G, that have a less distinct mapping of infall time GES orbital energy to the present-day orbital energy, in contrast to Au-5G and Au-15G in the right-most panels in Figure \ref{fig:postmerge}. The trend against radius is also more blurred but still present as seen in the center panel, such that lower [Fe/H] stars tend to originate from the outskirts of the progenitor at infall.


\bsp	
\label{lastpage}
\end{document}